\documentclass[12pt,journal,onecolumn]{IEEEtran}
\usepackage{epsfig,rotating,setspace,latexsym,amsmath,epsf,amssymb,amsfonts,bm,theorem,subfigure,epstopdf,graphicx}
\usepackage{cite,authblk}
\usepackage{algorithmic,algorithm}
\usepackage{setspace}
\usepackage{color}
\usepackage{dsfont}
 
\newtheorem{lemma}{Lemma}

\newenvironment{Proof}[1]{\medskip\par\noindent{\bf Proof:\,}\,#1}{{\mbox{\,$\blacksquare$}\par}}

\allowdisplaybreaks
 
\begin{document}
 
\title{Timestomping Vulnerability of Age-Sensitive Gossip Networks}
 
\author{Priyanka Kaswan \qquad Sennur Ulukus\\
        \normalsize Department of Electrical and Computer Engineering\\
        \normalsize University of Maryland, College Park, MD 20742\\
        \normalsize  \emph{pkaswan@umd.edu} \qquad \emph{ulukus@umd.edu}}
 
\maketitle

\vspace*{-2cm}

\begin{abstract}
We consider gossip networks consisting of a source that maintains the current version of a file, $n$ nodes that use asynchronous gossip mechanisms to disseminate fresh information in the network, and an oblivious adversary who infects the packets at a target node through data \emph{timestamp manipulation,} with the intent to replace circulation of fresh packets with outdated packets in the network. We demonstrate how network topology capacitates an adversary to influence age scaling in a network. We show that in a fully connected network, a single infected node increases the expected age from $O(\log n)$ to $O(n)$. Further, we show that the optimal behavior for an adversary is to reset the timestamps of all outgoing packets to the current time and of all incoming packets to an outdated time for the infected node; thereby preventing any fresh information to go into the infected node, and facilitating acceptance of stale information out of the infected node into other network nodes. Additionally, if the adversary allows the infected node to accept a small fraction of incoming packets from the network, then a large network can manage to curb the spread of stale files coming from the infected node and pull the network age back to $O(\log n)$. Lastly for fully connected network, we show that if an infected node contacts only a single node instead of all nodes of the network, the system age can still be degraded to $O(n)$. These show that fully connected nature of a network can be both a benefit and a detriment for information freshness; full connectivity, while enabling fast dissemination of information, also enables fast dissipation of adversarial inputs. We then analyze the unidirectional ring network, the other end of the network connectivity spectrum, where we show that the adversarial effect on age scaling of a node is limited by its distance from the adversary, and the age scaling for a large fraction of the network continues to be $O(\sqrt{n})$, unchanged from the case with no adversary. We finally support our findings with simulations.
\end{abstract}

\section{Introduction}

Sensor networks generally have limited resources, which prevents them from implementing traditional computer security techniques, making them vulnerable to adversarial attacks. Uncertain dynamics of such networks often force them to rely on decentralized \emph{gossip protocols} \cite{Demers1987EpidemicAF,Minsky02cornellthesis, vocking2000, Pittel1987OnSA, deb2006AlgebraicGossip, devavrat2006, Sanghavi2007GossipFileSplit, amazondynamo, Cassandra, Yates21gossip_traditional,Yates21gossip, baturalp21comm_struc, Bastopcu21gossip, kaswan22slicingcoding} for information dissemination, where information is exchanged between nodes repeatedly and asynchronously using their local status. Gossip protocols were  introduced and have been widely used in the context of distributed databases. In this work, we consider the presence of an adversary in a gossip network \cite{Nguyen17interferencegame, Garnaev19jamming, Xiao18jamming, kaswan22jammerring, Banerjee22adversary, Banerjee22game, Augustine16_adversaries, Georgiou08_complexitygossip}, who corrupts the gossip operation by manipulating the timestamps of some data packets flowing in the network, a technique known as \emph{timestomping} \cite{minnaard2014timestomping}, with the goal of bringing about staleness and inefficiency to the network. A timestomping attack can be launched in many ways. For instance, a malicious insider node can deviate from the gossip protocol and inject old packets by rebranding them as fresh packets via timestamp manipulation, while maintaining the gossiping frequency to evade suspicion. Other methods include \emph{meddler in the middle} (MITM) attacks, where the adversary inserts its node undetected between two nodes and manipulates communication, and \emph{eclipse} attacks  where the adversary manipulates the target node by redirecting its inbound and outbound links away from legitimate neighboring nodes to adversary controlled nodes, thereby isolating the node from the rest of the network, as encountered in gossip based blockchain networks.

Most prior works on gossip networks consider total dissemination time of a message in the network as the performance metric. For instance, \cite{vocking2000} shows that dissemination of a single rumor to $n$ nodes takes $O(\log n)$ rounds, \cite{deb2006AlgebraicGossip} shows that $n$ messages can be disseminated to $n$ nodes in $O(n)$ time in fully connected networks using random linear coding (RLC), \cite{devavrat2006} provides an analogous result for arbitrarily connected graphs, and  \cite{Sanghavi2007GossipFileSplit} analyzes dissemination of messages by dividing them into pieces. However, highly dynamic nature of data sources in modern applications prevents these networks from waiting for a specific message to reach all nodes of the network before fresh information can be circulated. Distributed databases \cite{amazondynamo, Cassandra}, for example, employ timestamp versioning, wherein every new information is created with a timestamp value taken from the system clock. When two nodes come in contact to exchange information, the timestamps of data at both nodes are compared and the node carrying the data with older timestamp discards its data for the fresher data of the other node. 

In this regard, age of information \cite{Kosta17agesurvey, Sun19agesurvey, yates21agesurvey} may be a more suitable indicator of network efficiency. Given $U_i(t)$ as the timestamp of the packet with node $i$ at time $t$, the instantaneous age of information is given by $X_i(t)=t-U_i(t)$. The nodes wish to have access to the most up-to-date information at all times, and therefore, are prompted to decrease $X_i(t)$ by fetching packets with more recent timestamps, e.g., with higher $U_i(t)$. Gossip networks have been studied from timeliness perspective in \cite{Yates21gossip_traditional, Yates21gossip, baturalp21comm_struc, Bastopcu21gossip, kaswan22slicingcoding}. \cite{Yates21gossip_traditional, Yates21gossip} derive the recursive age equations using stochastic hybrid system framework for age, \cite{baturalp21comm_struc} studies the expected version age in clustered gossip networks,  \cite{Bastopcu21gossip} extends these results to the binary freshness metric, \cite{kaswan22slicingcoding} considers age scaling in gossip networks using file slicing and network coding, and \cite{kaswan22jammerring} studies the effects of jamming adversaries on gossip age in ring networks. 

Timestomping is often used by malware authors as an anti-forensics technique to make files blend in with the rest of the system. In this work, an adversary uses timestomping with the goal of worsening the expected age in the network. Consider two nodes, $A$ and $B$, that randomly come in contact to exchange information and consider the presence of adversary at node $A$ capable of altering timestamps of all incoming and outgoing files. If node $A$ is outdated compared to node $B$, the adversary would be inclined to increase the timestamp of an outgoing packet from node $A$ to make it appear fresher so as to misguide node $B$ into discarding its packet in favor of a staler packet, and also, decrease the timestamp of an incoming packet from node $B$ so as to avoid its acceptance at node $A$. Conversely, if node $A$ is more up-to-date than node $B$, the adversary would reduce timestamps of outgoing packets and increase timestamps of incoming packets to make node $B$ reject fresher files and node $A$ accept staler files. More the manipulated timestamps digress away from their true value, higher are the chances of error in deciding which packet should be discarded, since this decision is based on a comparison of timestamps. At time $t$, the maximum error is caused when file timestamp is either changed to the current time $t$ or the earliest time $0$. Thus, we consider an adversary, who, for each packet, makes the decision of changing its timestamp to either $t$ or $0$. The adversary is \emph{oblivious} in that it does not look into a packet and see its actual timestamp. Thus, the adversary changes the timestamp to either $t$ or $0$ probabilistically.

In this paper, we consider gossip networks where an adversary captures a node and manipulates the timestamps of packets coming into and going out of the node, and demonstrate how network topology capacitates an adversary to influence age scaling in a network. We show that in a fully connected network (FCN), see Fig.~\ref{fig:fully_conn_net_eclipse_attack}, one infected node can single-handedly suppress the availability of fresh information in a large network of $n$ users employing a gossip protocol, and increase the expected age from $O(\log n)$ found in \cite{Yates21gossip_traditional} to $O(n)$. In addition, we show that the optimal action for the adversary is to always increase the timestamp of every outgoing packet to $t$ and decrease the timestamp of every incoming packet to $0$, in effect, preventing all incoming files from being accepted by the infected node and actively persuading other nodes to always accept outgoing packets from the infected node. Further, we show that if the the infected node is allowed to accept even a small fraction of incoming packets from the network, then a large network can curb the spread of stale files coming from the infected node by effectively lowering the infected node's age. These observations show how the fully connected nature of a network can be both a benefit and a detriment for network staleness. Additionally, we show that if the malicious node contacts only one other node instead of all nodes of the network, see Fig.~\ref{fig:fully_conn_net_MITM_attack}, the system age can still be degraded to $O(n)$, which highlights how little an effort is needed on the part of the adversary to bring down the freshness of the entire network.  

We then switch to the other end of the network connectivity spectrum and analyze the unidirectional ring network (URN), see Fig.~\ref{fig:unidirectional_ring_model}. As opposed to the fully connected network, where each node uniformly at random contacts all nodes of the network, in unidirectional ring each node devotes all its resources to establish contact with just one other node. Here we show that the adversarial effect on age scaling of a node is limited by its distance from the adversary, and the age scaling for a larger fraction of the network still continues to be $O(\sqrt{n})$ irrespective of adversarial policy. Note that in a unidirectional (or bidirectional) ring with no adversary, the age scaling of a node follows $O(\sqrt{n})$ for a network size $n$. We show that the optimal action for the adversary here again is to always increase the timestamp of every outgoing packet to $t$ and to decrease the timestamp of every incoming packet to $0$, similar to the fully connected network.

In the remainder of the paper, we frequently make use of the following four asymptotic notations: $O(\cdot)$, $o(\cdot)$, $\omega(\cdot)$ and $\Omega(\cdot)$ for big–O, little–o, big–Omega and little–omega, respectively. If $f(n)$ and $g(n)$ are functions that map positive integers to positive real numbers, then 
\begin{itemize}
    \item $f(n)=O(g(n))$ iff there exists a positive integer $n_0$ and a positive constant $c$ such that $f(n) \leq c g(n)$ for all $n\geq n_0$.
    \item $f(n)=o(g(n))$ iff for any positive constant $c$, there exists a positive integer $n_0$ such that $f(n) < c g(n)$ for all $n\geq n_0$, 
    i.e., $\lim_{n \to \infty}\frac{f(n)}{g(n)}=0$.
    \item $f(n)=\Omega(g(n))$ iff there exists a positive integer $n_0$ and a positive constant $c$ such that $f(n) \geq c g(n)$ for all $n\geq n_0$.
    \item $f(n)=\omega(g(n))$ iff for any positive constant $c$, there exists a positive integer $n_0$ such that $f(n) > c g(n)$ for all $n\geq n_0$, i.e., $\lim_{n \to \infty}\frac{f(n)}{g(n)}=\infty$.
\end{itemize}
For example, $f(n)=2n^2+3n$ is $O(n^2)$, $O(n^3)$, $o(n^3)$, $\Omega(n)$, $\Omega(n^2)$, $\Omega(1)$ and $\omega(1)$. For $f(\cdot)$ and $g(\cdot)$ to be of the same order, $f(n)$ should be both $O(g(n))$ and $\Omega(g(n))$, for instance, $f(n)$ is exactly order $n^2$, since constant multiples of $n^2$ serve as both lower and upper bounds to $f(n)$.

\section{System Model and SHS Characterization} \label{sec:sysmodel_fullyconn}

\begin{figure}[t]
\centerline{\includegraphics[width=0.6\linewidth]{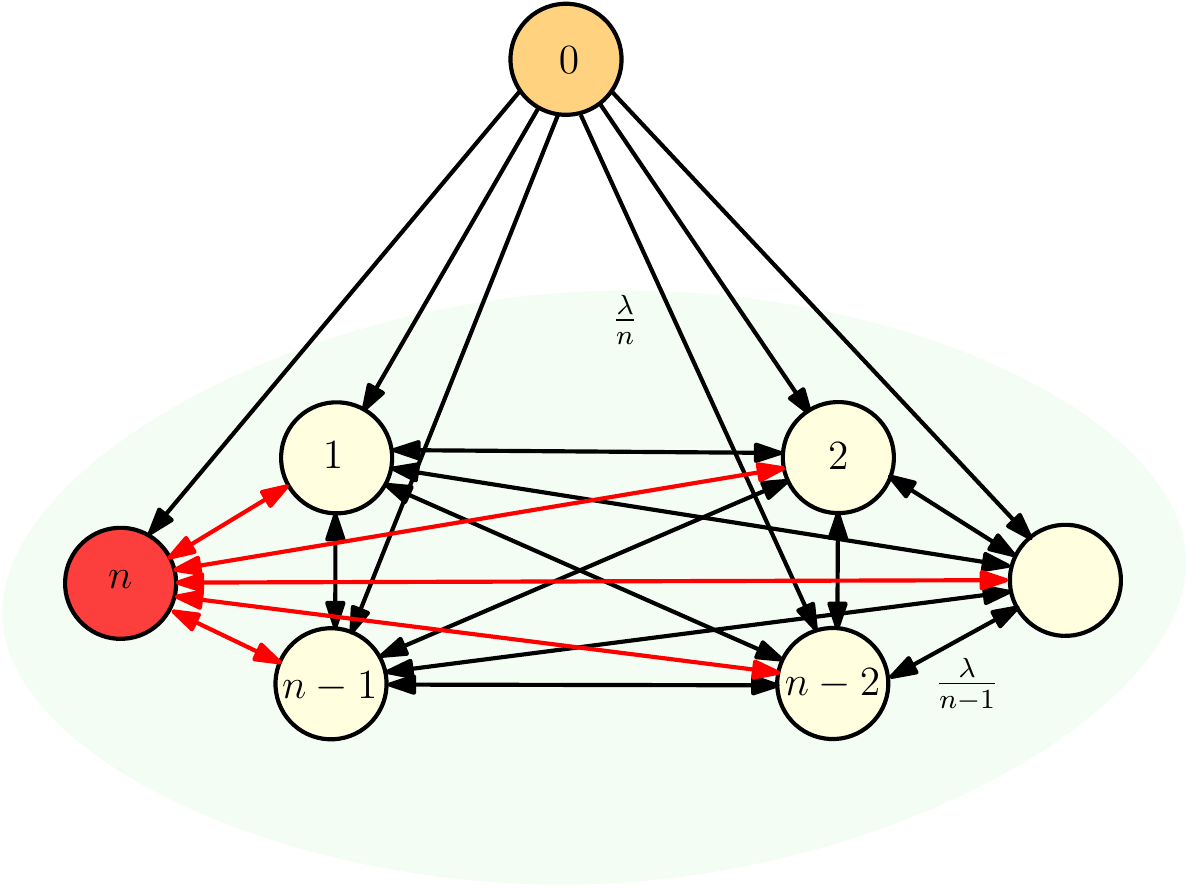}}
\caption{Fully connected network of $n$ nodes with an infected node.}
\label{fig:fully_conn_net_eclipse_attack}
\end{figure}

We first study a fully connected network (FCN), shown in Fig.~\ref{fig:fully_conn_net_eclipse_attack}, which comprises a source and $n$ user nodes $\mathcal{N}=\{1,\ldots,n\}$. The source, alternatively referred to as node $0$, is assumed to always posses the latest file packet and consequently has zero age at all times. The nodes wish to acquire the most up-to-date file to lower their average age from the source, who updates each user node as a Poisson process with rate $\frac{\lambda}{n}$. Further, a user node $i$ randomly sends its current packet to a user node $j$ according to a Poisson process with rate $\lambda_{ij}=\frac{\lambda}{n-1}$. Thus, all nodes send out updates after exponential inter-update times with a total rate $\lambda$. Let $U_j(t)$ denote the timestamp marked on the file stored at node $j$. Then, at the receiving node $j$, the claimed timestamp of the incoming packet is compared with $U_j(t)$ to determine which packet should be kept. Note that a node always accepts an update from the source which generates update packets with the current timestamp $t$. 

We assume that the highest index node, node $n$, is under attack by an adversary that manipulates the timestamps of all incoming and outgoing packets of node $n$. For the outgoing packets, the adversary chooses to increase the timestamp (to current time $t$) with probability $p$ and decrease the timestamp to $0$ with probability $1-p$. Similarly, the adversary increases and decreases the timestamp of incoming packets with probability $1-q$ and $q$, respectively. We will refer to the nodes in set $\mathcal{N}_R=\{1,\ldots,n-1\}$ as \emph{regular} nodes and node $n$ as the \emph{infected} node. We assume that the infected node always accepts packets from the source like other nodes, delivered to it with rate $\frac{\lambda}{n}$, which helps the adversary evade suspicion of malicious activity by maintaining a remote contemporary relevance of the contents of its manipulated packets.

We denote the long-term average age at node $i$ by $v_i$, where $v_i=\lim_{t \to \infty} \mathbb{E}[X_i(t)]$, and wish to study its extent of deterioration through timestomping. Note that the actual instantaneous age at node $i$ is $X_i(t)= t-\bar{U}_i(t)$, where  $\bar{U}_i(t)$ indicates the true packet generation time, which can be different from the claimed timestamp $U_i(t)$ if the file timestamp has been tampered with. For a set of nodes $S$ at time $t$, let $X_{N(S)}(t)$ indicate the actual instantaneous age of the node claiming to possess the most recent timestamped packet in set $S$, i.e., $X_{N(S)}(t)=X_{\arg \max_{j\in S} U_j(t)}(t)$. We define $v_S=\lim_{t \to \infty} \mathbb{E}[X_{N(S)}(t)]$. Here we would like to point out that in a network without adversary where all files are marked with true timestamps, $X_{N(S)}(t)$ reduces to $X_S(t)= \min_{j\in S}X_j(t)$ defined in \cite{Yates21gossip_traditional}, since the node with the highest timestamp will also have the lowest age in the set $S$. 

Reference \cite{Yates21gossip_traditional} demonstrates how stochastic hybrid system (SHS) models yield linear equations useful for deriving long-term average age at nodes in a gossip network of $n$ users with a given topology. Due to the presence of a timestomping adversary, we choose the continuous state for our SHS model as $(\pmb{X}(t),\pmb{U}(t))\in \mathbb{R}^{2n}$, where $\pmb{X}(t)=[X_1(t),\ldots,X_n(t)]$ denotes the instantaneous ages at the $n$ nodes and $\pmb{U}(t)=[U_1(t),\ldots,U_n(t)]$ denotes the timestamps marked on the packets at the $n$ nodes at time $t$. The convenience of the SHS based age characterization follows from the presence of a single discrete mode with trivial stochastic differential equation $(\pmb{\dot X}(t),\pmb{\dot U}(t))=(\pmb{1}_n,\pmb{0}_n)$, where the age at each node grows at unit rate when there is no update transfer, since the timestamps of the node packets do not change between such transitions. Consider a test function $\psi:\mathbb{R}^{2n}\times [0,\infty) \to \mathbb{R}$ that is time-invariant, i.e., its partial derivative with respect to $t$ is $\frac{\partial\psi(\pmb{X},\pmb{U},t)}{\partial t}=0$, such that we are interested in finding its long-term expected value $\mathbb{E}[\psi]=\lim_{t \to \infty} \mathbb{E}[\psi(\pmb{X}(t),\pmb{U}(t),t)]$. Since the test function only depends on the continuous state values $(\pmb{X},\pmb{U})$ and is time-invariant, for simplicity, we will drop the third input $t$ and write $\psi(\pmb{X},\pmb{U},t)$ as $\psi(\pmb{X},\pmb{U})$, which we assume to flow according to the differential equation $\dot \psi(\pmb{X}(t),\pmb{U}(t))=1$. Let $\mathcal{L}$ correspond to the set of directed edges $(i,j)$, such that node $i$ sends updates to node $j$ on this edge according to a Poisson process of rate $\lambda_{ij}$, with this transition resetting the state $(\pmb{X},\pmb{U})$  at time $t$ to $\phi_{i,j}(\pmb{X},\pmb{U},t)\in \mathbb{R}^{2n}$ post transition. Defining $\mathbb{E}[\psi(\phi_{i,j})]=\lim_{t \to \infty} \mathbb{E}[\psi(\phi_{i,j}(\pmb{X}(t),\pmb{U}(t),t))]$, \cite[Thm.~1]{hespanhashs} yields 
\begin{align} \label{eqn:hespanha_eqn}
    0=1+\sum_{(i,j)\in \mathcal{L}}\lambda_{ij}(\mathbb{E}[\psi(\phi_{i,j})]- \mathbb{E}[\psi] )
\end{align}
which is similar to derivations in \cite{Yates21gossip_traditional}, where the left side becomes $0$ as expectations stabilize. We will use this equation repeatedly by defining a series of time-invariant test functions appropriate for our analysis. For more details, the reader is encouraged to consult references \cite{hespanhashs} and \cite{Yates21gossip_traditional}.

\section{Fully Connected Network} \label{sect:no-capture}

Note that packets arriving at infected node $n$ from a node $i\in \mathcal{N}_R$ with rate $\frac{\lambda}{n-1}$ Poisson process are accepted (or discarded) with probability $1-q$ (or $q$) when the adversary changes timestamp of incoming packet to $t$ (or $0$) to make it appear fresh (or stale). This is equivalent to packets arriving at node $n$ from node $i$ with thinned Poisson process with rate $\lambda_{in}=\frac{(1-q)\lambda}{n-1}$ such that these packets are always accepted. The remaining packets are always discarded and have no effect on age dynamics of the system. Similarly, as the outgoing packets from the infected node $n$ are accepted at node $i\in \mathcal{N}_R$ with probability $p$, this is equivalent to node $n$ sending packets with timestamp $t$ to node $i$ with a thinned Poisson process of rate $\lambda_{ni}=\frac{p\lambda}{n-1}$ such that these packets are always accepted. The rates $\lambda_{ij}$ for each transition $(i,j)$ are given as
\begin{align} 
\lambda_{ij} = \begin{cases} 
\frac{\lambda}{n}, & i=0,j\in \mathcal{N}\\
\frac{\lambda}{n-1}, & i,j\in \mathcal{N}_R\\
\frac{p\lambda}{n-1}, & i=n,j\in \mathcal{N}_R\\
\frac{(1-q)\lambda}{n-1}, & i\in \mathcal{N}_R,j=n
\end{cases}
\end{align}

Thus, based on transition $(i,j)$ at time $t$, the reset map $\phi_{i,j}(\pmb{X},\pmb{U},t)=[X_1',\ldots,X_n',U_1',\ldots,U_n']$ can be described by
\begin{align}\label{eqn:eclipse_U_resetmap}
U_{\ell}' = \begin{cases} 
t, & i=0,j\in \mathcal{N},\ell=j\\
\max\{U_i,U_{\ell}\}, & i,j\in \mathcal{N}_R,\ell=j\\
t, & i=n,j\in \mathcal{N}_R,\ell=j\\
t, & i\in \mathcal{N}_R,j=n,\ell=j\\
U_{\ell}, & \text{otherwise}
\end{cases}
\end{align}
and
\begin{align} \label{eqn:X_reset}
X_{\ell}' = \begin{cases} 
0, & i=0,j\in \mathcal{N},\ell=j\\
X_{N(\{i,\ell\})}, & i,j\in \mathcal{N}_R,\ell=j\\
X_n, & i=n,j\in \mathcal{N}_R,\ell=j\\
X_i, & i\in \mathcal{N}_R,j=n,\ell=j\\
X_{\ell}, & \text{otherwise}
\end{cases}
\end{align}

Here, $X_{N(S)}= X_{\arg \max_{j\in S}U_j}$ for state $(\pmb{X},\pmb{U})$ and a subset of nodes $S$. Since all regular nodes have statistically similar age processes, every arbitrary set $S_k$ of $k$ regular nodes will have the same expected age $v_{S_k}$, $S_k \subseteq \mathcal{N}_R$, with $v_{S_1}=v_1$. We pick our first test function to be $\psi(\pmb{X},\pmb{U})=X_{N(S_k)}$, which is modified upon transition $(i,j)$ to  $\psi(\phi_{i,j}(\pmb{X},\pmb{U},t))=X_{N(S_k)}'$. This in turn is characterized using (\ref{eqn:eclipse_U_resetmap}) and (\ref{eqn:X_reset}) as 
\begin{align} \label{eqn:X_NS_k_reset}
X_{N(S_k)}' = \begin{cases} 
0, & i=0,j\in S_k,\ell=j\\
X_{N(S_{k}\cup\{i\})}, & i\in \mathcal{N}_R\backslash S_k,j\in S_k,\ell=j\\
X_n, & i=n,j\in S_k,\ell=j\\
X_{N(S_k)}, & \text{otherwise}
\end{cases}
\end{align}

To check if this test function satisfies $\dot \psi(\pmb{X}(t),\pmb{U}(t))=1$, remember that $\pmb{\dot X}(t)=\pmb{1}_n$ and hence $X_{N(S_k)}(t)=X_{\arg \max_{j\in S_k} U_j(t)}(t)$ grows at unit rate between transitions. Employing (\ref{eqn:hespanha_eqn}), this test function yields,
\begin{align}\label{eqn:eclipse_attack_hespanha_equation}
    0=&1+\frac{k\lambda}{n}(0-v_{S_k})+\frac{(n-k-1)k\lambda}{n-1}(v_{S_{k+1}}-v_{S_k})+\frac{kp\lambda}{n-1}(v_{n}-v_{S_k}) 
\end{align}
which upon rearrangment gives
\begin{align}\label{eqn:eclipse_vSk}
    v_{S_k}=\frac{\frac{1}{k\lambda}+\frac{n-k-1}{n-1}v_{S_{k+1}}+\frac{pv_n}{n-1}}{\frac{1}{n}+\frac{n-k-1}{n-1}+\frac{p}{n-1}}
\end{align}

Our second test function is simply $\psi(\pmb{X},\pmb{U})=X_n$, i.e., the age at infected node, such that its $(i,j)$ transition map is 
\begin{align} \label{eqn:X_n_reset}
X_n' = \begin{cases} 
0, & i=0,j=n,\ell=j\\
X_i, & i\in \mathcal{N}_R,j=n,\ell=j\\
X_n, & \text{otherwise}
\end{cases}
\end{align}
which upon proceeding similarly to (\ref{eqn:eclipse_attack_hespanha_equation}) and (\ref{eqn:eclipse_vSk}), gives
\begin{align}\label{eqn:Xn_eclipse_adversary}
    v_n=\frac{\frac{1}{\lambda}+(1-q)v_1}{\frac{1}{n}+(1-q)}
\end{align}

Our goal is to obtain an analytical expression for the expected age of a regular node $v_{S_1}=v_1$, by making use of (\ref{eqn:eclipse_vSk}) and (\ref{eqn:Xn_eclipse_adversary}). We now proceed to analyze how the choice of probabilities $p$ and $q$ by the adversary affects the expected age at regular nodes in the regime of large $n$.

\subsection{FCN Case~1: $p>0,q=1$} \label{case: fullyconnec_case1}

These probabilities pertain to the case where the infected node blocks all incoming packets from the regular nodes, and at the same time sends out packets with timestamps manipulated to $t$ with rate $\frac{p\lambda}{n-1}$ to each regular node, misleading them into accepting these packets. 

Using $\frac{1}{n} \le \frac{1}{n-1} $ and $p\leq1$ in the denominator of (\ref{eqn:eclipse_vSk}) gives the following lower bound
\begin{align} \label{eqn:case1_v_k_lowerbound_fullyconn}
    v_{S_k} \geq \frac{\frac{1}{k\lambda}+\frac{n-k-1}{n-1}v_{S_{k+1}}+\frac{pv_n}{n-1}}{\frac{2}{n-1}+\frac{n-k-1}{n-1}}
\end{align}
Letting $y_k=v_{S_k}\frac{n-k}{n-1}$ in (\ref{eqn:case1_v_k_lowerbound_fullyconn}) gives 
\begin{align}
    y_k \geq \frac{n-k}{n-k+1}\bigg( y_{k+1}+ \frac{1}{k\lambda} +  \frac{pv_n}{n-1} \bigg)
\end{align}
Starting from $y_1=v_1$, and successively substituting for $y_2, y_3, \ldots, y_{n-1}$, we obtain
\begin{align}\label{eqn:case1_v_1_lowerbound_fullyconn}
    v_1\geq& \frac{1}{\lambda} \sum_{k=1}^{n-1} \frac{n-k}{nk} +pv_n\sum_{k=1}^{n-1}\frac{n-k}{n(n-1)} \nonumber \\
    =& \frac{1}{\lambda} \sum_{k=1}^{n-1}\frac{1}{k} -\frac{1}{\lambda}\frac{n-1}{n} +\frac{pv_n}{n(n-1)}\sum_{k=1}^{n-1} k \nonumber \\
    =& \Omega(\log n) + \frac{pv_n}{2} 
\end{align}
since $\sum_{k=1}^{n-1}\frac{1}{k}$ grows asymptotically as $\log n$ and $0\leq \frac{n-1}{n} \leq 1$. The $v_n$ in the second term can in turn be obtained by substituting $q=1$ in (\ref{eqn:Xn_eclipse_adversary}), giving $v_n=\frac{n}{\lambda}$.
Hence, 
\begin{align} \label{eqn:v1_vn_by_2_case1}
    v_1\geq \Omega(\log n) + \frac{pn}{2\lambda}=\Omega(n)
\end{align}
In other words, the age at a regular node in this case scales at least as $n$. In fact, we next show that the age at a regular node in this case scales at most as $n$ as well, i.e., $v_1$ is $O(n)$. 

We add $\frac{1}{n-1}-\frac{p}{n-1}-\frac{1}{n}=\frac{1-np}{n(n-1)}$ (which is $\leq 0 $ for large $n$) to the denominator of (\ref{eqn:eclipse_vSk}) to get the following upper bound 
\begin{align} \label{eqn:case1_v_1_upperbound_fullyconn}
    v_{S_k} \leq \frac{\frac{1}{k\lambda}+\frac{n-k-1}{n-1}v_{S_{k+1}}+\frac{pv_n}{n-1}}{\frac{1}{n-1}+\frac{n-k-1}{n-1}}
\end{align}
where we again employ $y_k=v_{S_k}\frac{n-k}{n-1}$ which gives 
\begin{align}
y_k \leq y_{k+1}+\frac{1}{k\lambda}+\frac{pv_n}{n-1}  
\end{align}
Here an iterative substitution for $y_2, y_3, \ldots, y_{n-1}$ with $y_1=v_1$ yields 
\begin{align}\label{eqn:case1_v_1_finalupperbound_fullyconn}
    v_1\leq \frac{1}{\lambda} \sum_{k=1}^{n-1}\frac{1}{k} + pv_n
\end{align}
Since $\sum_{k=1}^{n-1}\frac{1}{k}=O(\log n)$ and $v_n=\frac{n}{\lambda}=O(n)$, we conclude $v_1$ is $O(n)$.

To put this deterioration in age scaling into perspective, remember that in a fully disconnected network with no gossiping \cite{baturalp21comm_struc}, expected age at each node also scales as $O(n)$, to be exact, $\frac{n}{\lambda}$, a fact that will come handy later.  

\subsection{FCN Case~2: $p=0,q\leq 1$} \label{case: fullyconnec_case2}

In this case, the infected node accepts timestomped packets from the $n-1$ regular nodes with rate $\frac{(1-q)\lambda}{n-1}$ but effectively does not transmit any files, even when it possesses the latest file, thereby limiting its contribution, positive or negative, to the system age. 

Substituting $p=0$ in (\ref{eqn:eclipse_vSk}) gives
\begin{align}
    v_{S_k}=\frac{\frac{1}{k\lambda}+\frac{n-k-1}{n-1}v_{S_{k+1}}}{\frac{1}{n}+\frac{n-k-1}{n-1}}
\end{align}
Again defining $y_k=v_{S_k}\frac{n-k}{n-1}$, using $\frac{n-k-1}{n-1} \geq \frac{n-k-1}{n}$ in the denominator gives $y_k \leq \frac{n}{n-1}(y_{k+1}+\frac{1}{k\lambda})$. Since $\left(\frac{n}{n-1}\right)^{j} \leq \left(\frac{n}{n-1}\right)^{n-1}$ for $j \leq n-1$, an iterative substitution for $k=\{1,\ldots,n-1\}$ gives
\begin{align}\label{eqn:case2_v_1_upperbound_fullyconc} 
    v_1=y_1 \leq  \left(\frac{n}{n-1}\right)^{n-1}  \left( \frac{1}{\lambda} \sum_{k=1}^{n-1}\frac{1}{k} \right) = O(\log n)
\end{align}
where we employed $\lim _{n \to \infty} \left(\frac{n}{n-1}\right)^{n-1} = e$ and $ \sum_{k=1}^{n-1}\frac{1}{k}=O(\log n)$. Hence, such an adversary does not affect the network age scaling. Further, specifically for the case of $q<1$, using $\frac{1}{n} \geq 0$ in (\ref{eqn:Xn_eclipse_adversary}) gives 
\begin{align}\label{eqn:eclipse_vn_case2}
    v_n=\frac{\frac{1}{\lambda}+(1-q)v_1}{\frac{1}{n}+(1-q)} \leq \frac{1}{\lambda (1-q)}+ v_1
\end{align}

Hence $v_n$ scales as $O(\log n)$ similar to $v_1$, which indicates that the deterioration of age at the infected node itself is negligible as long as it accepts even a small fraction of incoming packets. For the case of $q=1$, when infected node does not accept any packets from the gossip process, (\ref{eqn:Xn_eclipse_adversary}) gives $v_n=O(n)$, since it behaves as an isolated node.

\subsection{FCN Case~3: $p>0,q<1$} \label{case: fullyconnec_case3}

In this case, the adversary partially both accepts and sends timestomped packets to regular nodes. Here, we employ a combination of (\ref{eqn:case1_v_1_finalupperbound_fullyconn}) and (\ref{eqn:eclipse_vn_case2}) to get
\begin{align}
    v_1\leq& \frac{1}{\lambda} \sum_{k=1}^{n-1}\frac{1}{k} + p\left( \frac{1}{\lambda (1-q)}+ v_1 \right)
\end{align}
which upon rearrangement gives
\begin{align}\label{eqn:case3_v_1_upperbound_fullyconc}
    v_1\leq \frac{1}{(1-p)}\left(\frac{1}{\lambda} \sum_{k=1}^{n-1}\frac{1}{k} + \frac{p}{\lambda (1-q)}\right)
\end{align}
where $\sum_{k=1}^{n-1}\frac{1}{k}=O(\log n)$. Hence, $v_1$ scales as $O(\log n)$ and (\ref{eqn:eclipse_vn_case2}) provides that $v_n$ also scales as $O(\log n)$. This case, in combination with \emph{Case~2}, highlights how the infected node fails to deteriorate the age scaling at regular nodes from $O(\log n)$ as long as the infected node receives packets from the gossip network.

Before we end this section, we remark that the maximum age deterioration in the network caused by the adversary happens when $p=1$ and $q=1$, i.e., a special case of \emph{Case~1}. To see this, we first note that $O(n)$ age is achieved only when $q=1$, as can be seen by comparing all the three cases above. Thus, we take $q=1$. When $q=1$, we have $v_n=\frac{n}{\lambda}$. To show the optimality of $p=1$, we rewrite (\ref{eqn:eclipse_vSk}) as follows where we use $v_n$ and $\frac{n}{\lambda}$ interchangeably
\begin{align}
    v_{S_k}=&\frac{\frac{1}{k\lambda}+\frac{n-k-1}{n-1}v_{S_{k+1}}+\frac{pv_n}{n-1}}{\frac{1}{n}+\frac{n-k-1}{n-1}+\frac{p}{n-1}} \nonumber\\
    =&\frac{\left(1-1+\frac{1}{k}\right)\frac{1}{\lambda}+\frac{n-k-1}{n-1}(v_n-v_n+v_{S_{k+1}})+\frac{pv_n}{n-1}}{\frac{1}{n}+\frac{n-k-1}{n-1}+\frac{p}{n-1}} \nonumber \\
    =& \frac{n}{\lambda}\left(\frac{\frac{1}{n}+\frac{n-k-1}{n-1}+\frac{p}{n-1}}{\frac{1}{n}+\frac{n-k-1}{n-1}+\frac{p}{n-1}}\right)+ \frac{\left(-1+\frac{1}{k}\right)\frac{1}{\lambda}+\frac{n-k-1}{n-1}(-v_n+v_{S_{k+1}})}{\frac{1}{n}+\frac{n-k-1}{n-1}+\frac{p}{n-1}} \nonumber \\
    =& \frac{n}{\lambda}- \frac{\left(1-\frac{1}{k}\right)\frac{1}{\lambda}+\frac{n-k-1}{n-1}(v_n-v_{S_{k+1}})}{\frac{1}{n}+\frac{n-k-1}{n-1}+\frac{p}{n-1}}
\end{align}    
Bringing $v_n=\frac{n}{\lambda}$ to the left hand side, we obtain
\begin{align}\label{eqn:vn-vsk}
    v_n-v_{S_k}=\frac{\left(1-\frac{1}{k}\right)\frac{1}{\lambda}+\frac{n-k-1}{n-1}(v_n-v_{S_{k+1}})}{\frac{1}{n}+\frac{n-k-1}{n-1}+\frac{p}{n-1}}
\end{align}
Substituting $k=n-1$ yields
\begin{align}
    v_n-v_{S_{n-1}}=\frac{\left(1-\frac{1}{n-1}\right)\frac{1}{\lambda}}{\frac{1}{n}+\frac{p}{n-1}}
\end{align}
Thus, $v_n-v_{S_{n-1}}$ is positive and a decreasing function of $p$. Hence, $v_{S_{n-1}}$ is an increasing function of $p$. Thereafter, using (\ref{eqn:vn-vsk}), we recursively show that $v_n-v_{S_{k}}$, for $k=n-2, n-3, \ldots, 1$, is a decreasing function of $p$, hence, $v_{S_{k}}$ for all $k$, is an increasing function of $p$. Therefore, $v_1=v_{S_{1}}$ is an increasing function of $p$. Hence, the age deterioration in the network caused by the adversary increases with increasing $p$, and the maximum happens when $p=1$.

\begin{figure}[t]
\centerline{\includegraphics[width=0.6\linewidth]{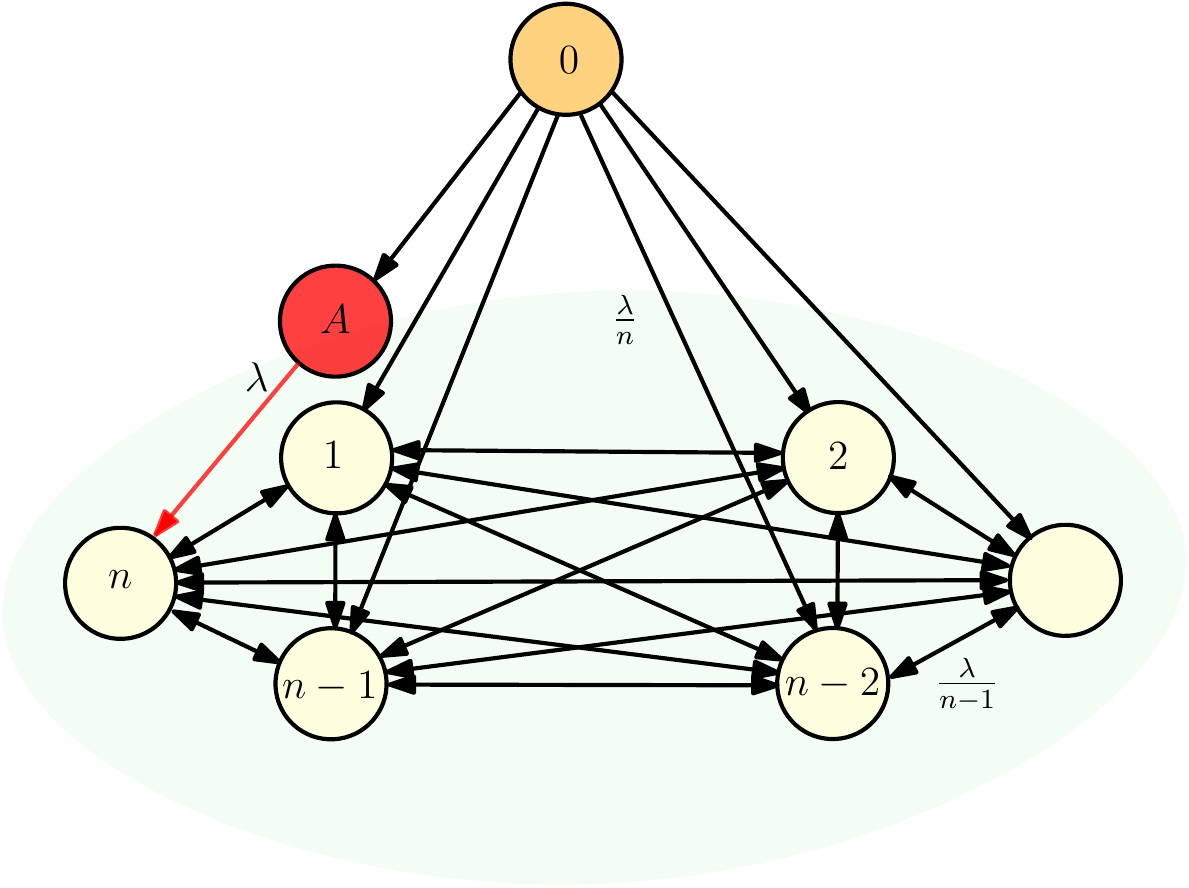}}
\caption{MITM attack on fully connected network of $n$ nodes.}
\label{fig:fully_conn_net_MITM_attack}
\end{figure}

\section{MITM Attack on Fully Connected Network} \label{sect:mitmattack}

In previous sections, the adversarial node was in direct contact with all other nodes due to the fully connected nature of the network, and the adversary could raise the system age to $O(n)$ using probabilities $p=1$ and $q=1$, effectively sending out all packets with timestamp $t$ and not accepting any packets from the gossip process. Here, an interesting question to ask is, if the network could do better if the adversary instead had access to only one node. To this end, we consider the network model of Fig.~\ref{fig:fully_conn_net_MITM_attack}, where the adversary, which we will refer to as node $A$, intercepts the updates to node $n$ coming from the source. In turn, the adversary sends updates with its total rate $\lambda$, after changing the timestamps of every outgoing packet to the current time, only to node $n$. 

Clearly the expected age at the adversary, denoted by $v_A$, scales as $O(n)$ since it is isolated from the gossip network and only receives updates from the source with rate $\frac{\lambda}{n}$. The two reset maps useful for our analysis are 
\begin{align} \label{eqn:mitm_X_NSkn_reset}
X_{N(S_k\cup\{n\})}' = \begin{cases} 
0, & i=0,j\in S_k,\ell=j\\
X_{N(S_{k+1}\cup\{n\})}, & i\in \mathcal{N}_R\backslash S_k,j\in S_k\cup\{n\},\ell=j\\
X_A, & i=A,j=n,\ell=j\\
X_{N(S_k\cup\{n\})}, & \text{otherwise}
\end{cases}
\end{align}
and
\begin{align} \label{eqn:mitm_X_NSk_reset}
X_{N(S_k)}' = \begin{cases} 
0, & i=0,j\in S_k,\ell=j\\
X_{N(S_{k+1})}, & i\in \mathcal{N}_R\backslash S_k,j\in S_k,\ell=j\\
X_{N(S_k\cup\{n\})}, & i=n,j\in S_k,\ell=j\\
X_{N(S_k)}, & \text{otherwise}
\end{cases}
\end{align}

We claim $v_{S_k\cup\{n\}} \geq \frac{v_A}{2}$, a loose lower bound that is trivially verified with induction as follows. Invoking (\ref{eqn:hespanha_eqn}) regarding (\ref{eqn:mitm_X_NSkn_reset}) for $k=n-1$ results in 
\begin{align}
    v_{S_{n-1}\cup\{n\}}=\frac{\frac{1}{\lambda}+  v_A}{\frac{n-1}{n} + 1} \geq O(1)+\frac{v_A}{2} \geq \frac{v_A}{2}
\end{align}
which verifies the claim for $k=n-1$. Next, we assume that the claim holds for $k+1$, i.e., $v_{S_{k+1}\cup\{n\}}\geq \frac{v_A}{2}$, and verify for $k$. Invoking (\ref{eqn:hespanha_eqn}) regarding (\ref{eqn:mitm_X_NSkn_reset}) for $k\leq n-2$ and using $\frac{1}{\lambda}>0$ in the numerator and $\frac{k}{n}\leq 1$ in the denominator gives 
\begin{align} \label{eqn:nmitm_vskn_lb_vA_by_2}
    \!\!\!v_{S_k\cup\{n\}}=& \frac{\frac{1}{\lambda}+ \frac{(k+1)(n-1-k)}{n-1} v_{S_{k+1}\cup\{n\}} + v_A}{\frac{k}{n} +\frac{(k+1)(n-1-k)}{n-1} + 1}  \nonumber\\
    \geq & \frac{\frac{(k+1)(n-1-k)}{n-1} v_{S_{k+1}\cup\{n\}}}{\frac{(k+1)(n-1-k)}{n-1} + 2} + \frac{v_A}{\frac{(k+1)(n-1-k)}{n-1} + 2}  \nonumber\\
    \geq & \frac{\frac{(k+1)(n-1-k)}{n-1} \frac{v_A}{2}}{\frac{(k+1)(n-1-k)}{n-1} + 2} + \frac{\frac{2v_A}{2}}{\frac{(k+1)(n-1-k)}{n-1} + 2} \nonumber\\
    =& \frac{v_A}{2}
\end{align}
Finally, we re-invoke (\ref{eqn:hespanha_eqn}) for (\ref{eqn:mitm_X_NSk_reset}) which results in 
\begin{align} \label{eqn:mitm_vSk_shs_stable_eqn}
    v_{S_k}=\frac{\frac{1}{k\lambda}+\frac{n-k-1}{n-1}v_{S_{k+1}}+\frac{v_{S_k\cup\{n\}}}{n-1}}{\frac{1}{n}+\frac{n-k-1}{n-1}+\frac{1}{n-1}}
\end{align}
Let $y_k=v_{S_k}\frac{n-k}{n-1}$, using $\frac{1}{n-1}\approx \frac{1}{n}$ for large $n$, (\ref{eqn:mitm_vSk_shs_stable_eqn}) becomes
\begin{align}
    y_k=&\frac{n-k}{n-k+1}\bigg( y_{k+1}+ \frac{1}{k\lambda} +  \frac{v_{S_k\cup\{n\}}}{n-1} \bigg) \nonumber \\
    \geq&\frac{n-k}{n-k+1} y_{k+1}+ \frac{(n-k)v_{S_k\cup\{n\}}}{(n-k+1)(n-1)} 
\end{align}
Starting from $y_1=v_1$, we substitute for $y_2, y_3, \ldots, y_{n-1}$ and use $v_{S_k\cup\{n\}} \geq \frac{v_A}{2}$ to obtain
\begin{align} 
    v_1\geq& \frac{1}{n(n-1)}\sum_{k=1}^{n-1} (n-k)v_{S_k\cup\{n\}} \nonumber \\
    \geq & \frac{v_A}{2n(n-1)}\sum_{k=1}^{n-1}(n-k) =\frac{v_A}{4} \label{eqn:mitm_v1_lb_va_by_4}
\end{align}

Hence, $v_1$ scales at least as $O(n)$ for all regular nodes. This result is far from intuitive, for it brings home the point how an adversary, with so little an effort as sending tampered packets to just one node, can bring down the freshness of an entire large gossip network.

\section{Unidirectinal Ring Network} \label{sect:unidirectionalring}

\begin{figure}[t]
\centerline{\includegraphics[width=0.6\linewidth]{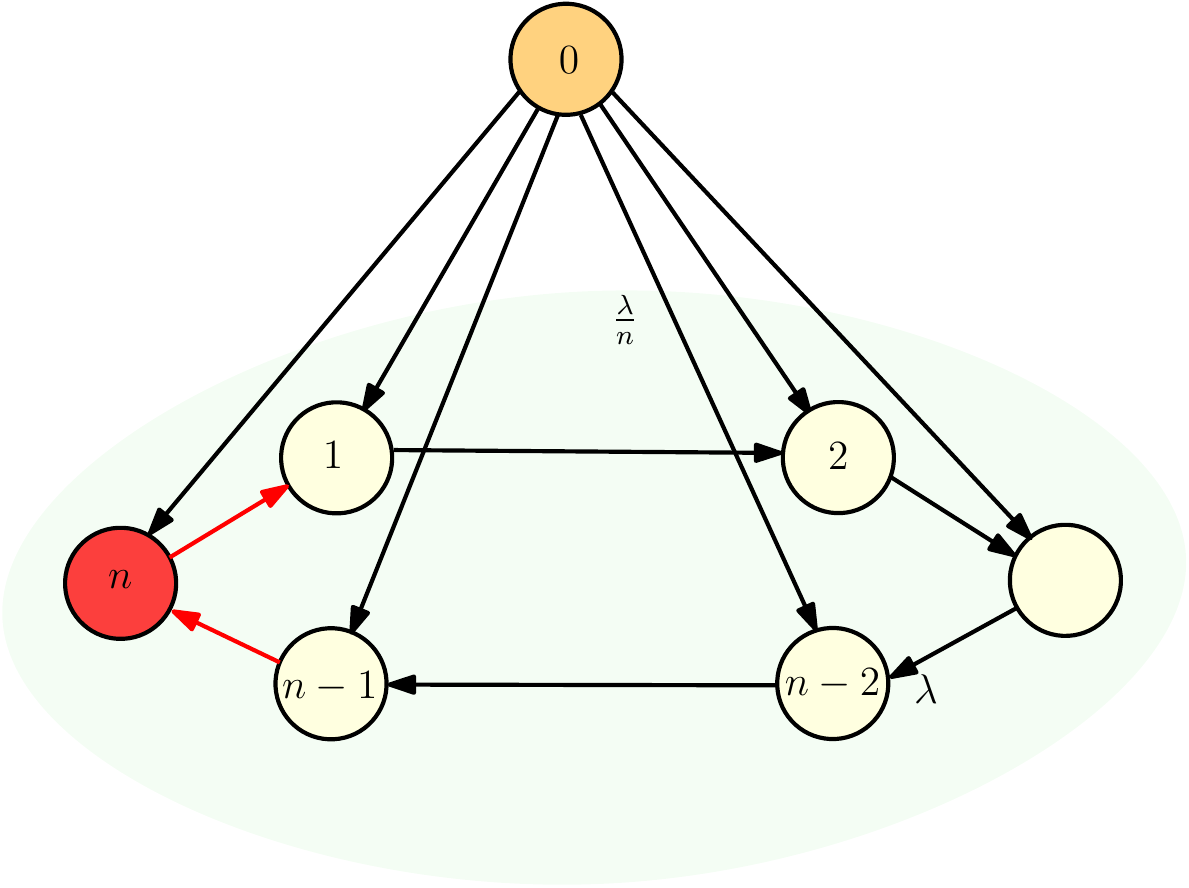}}
\caption{Unidirectional ring comprising $n-1$ regular nodes and an infected node.}
\label{fig:unidirectional_ring_model}
\end{figure}

We now consider the resource constrained unidirectional ring network (URN), shown in Fig.~\ref{fig:unidirectional_ring_model}, which represents the other end of the network connectivity spectrum. This network differs from the fully connected network in the sense that each node uses its entire rate $\lambda$ to transmit packets to a single node, i.e., node $i$ only transmits to node $j=(i \!\!\mod n)+1$. The source node, referred to as node $0$, always possesses the latest file packet, which it transmits to each user node in the set $\mathcal{N}=\{1,\ldots,n\}$ according to a Poisson process with rate $\frac{\lambda}{n}$. The nodes accept all packets coming from the source since it generates update packets with current timestamp $t$. We refer to node $n$ as the \emph{infected} node and nodes in the set $\mathcal{N}_R=\{1,\ldots,n-1\}$ as \emph{regular} nodes. Similar to Section~\ref{sec:sysmodel_fullyconn}, the oblivious adversary increases timestamps to current time $t$ of outgoing and incoming packets with probabilities $p$ and $1-q$, respectively. Effectively this means that packets arrive at the infected node $n$ from node $n-1$ according to a thinned Poisson process with rate $(1-q)\lambda$ while the infected node $n$ sends packets to node $1$ according to a thinned Poisson process with rate $p\lambda$. The rates $\lambda_{ij}$ for each transition $(i,j)$ are given as 
\begin{align} \label{eqn:lambda_ij_unidirectionalring}
\lambda_{ij} = \begin{cases} 
\frac{\lambda}{n}, & i=0,j\in \mathcal{N}\\
\lambda, & i\in \mathcal{N}_R\backslash\{n-1\},j=i+1\\
p\lambda, & i=n,j=1\\
(1-q)\lambda, & i=n-1,j=n
\end{cases}
\end{align}

Thus, based on transition $(i,j)$ at time $t$, the reset map $\phi_{i,j}(\pmb{X},\pmb{U},t)=[X_1',\ldots,X_n',U_1',\ldots,U_n']$ can be described by
\begin{align}
U_{\ell}' = \begin{cases} 
t, & i=0,j\in \mathcal{N},\ell=j\\
\max\{U_i,U_{\ell}\}, & i\in \mathcal{N}_R\backslash\{n-1\},j=i+1,\ell=j\\
t, & i=n,j=1,\ell=j\\
t, & i=n-1,j=n,\ell=j\\
U_{\ell}, & \text{otherwise}
\end{cases}
\end{align}
and
\begin{align} 
X_{\ell}' = \begin{cases} 
0, & i=0,j\in \mathcal{N},\ell=j\\
X_{N(\{\ell-1,\ell\})}, & i\in \mathcal{N}_R\backslash\{n-1\},j=i+1,\ell=j\\
X_n, & i=n,j=1,\ell=j\\
X_{n-1}, & i=n-1,j=n,\ell=j\\
X_{\ell}, & \text{otherwise}
\end{cases}
\end{align}

Keeping in mind that for state $(\pmb{X},\pmb{U})$ and subset $S=\{m-k+1,\ldots,m\}$, $X_{N(S)}= X_{\arg \max_{j\in S}U_j}$ chooses the age of the node claiming to possess the packet with the latest timestamp among the set of $k$ contiguous nodes, we wish to find the long-term average age $v_m$ at node $m$, where $v_m=\lim_{t \to \infty} \mathbb{E}[X_m(t)]$. To this purpose, for some $k \in \{1,\ldots,m-1\}$, the following test function will be useful
\begin{align} \label{eqn:X_N(m-k+1,ldots,m)unidirectionalring}
X_{N(\{m-k+1,\ldots,m\})}' = \begin{cases} 
0, & i=0,j\in S_k,\ell=j\\
X_{N(\{m-k,\ldots,m\})}, & i=m-k,j=m-k+1,\ell=j\\
X_{N(\{m-k+1,\ldots,m\})}, & \text{otherwise}
\end{cases}
\end{align}

Using (\ref{eqn:hespanha_eqn}), (\ref{eqn:lambda_ij_unidirectionalring}) and (\ref{eqn:X_N(m-k+1,ldots,m)unidirectionalring}), we get the following recursive equation
\begin{align}\label{eqn:v_m-k+1,ldots,m_unidirectional ring}
    v_{\{m-k+1,\ldots,m\}}&=\frac{1+\lambda v_{\{m-k,\ldots,m\}}}{\lambda+\frac{k\lambda}{n}} \nonumber \\
    &=\frac{\frac{1}{\lambda}}{1+\frac{k}{n}} +\frac{ v_{\{m-k,\ldots,m\}}}{1+\frac{k}{n}}
\end{align} 

Note that when $k=1$, $v_{\{m-k+1,\ldots,m\}} = v_{\{m,\ldots,m\}}=v_m$, since $X_{N(\{m\})}=X_m$. Solving together the set of equations obtained by substituting in (\ref{eqn:v_m-k+1,ldots,m_unidirectional ring}) all values of $k \in \{1,\ldots,m-1\}$, we obtain  
\begin{align}\label{eqn:v_m_obtained_recursively_unidirectionalring}
    v_m= \frac{1}{\lambda} \sum_{j=1}^{m-1} \prod_{k=1}^{j} \frac{1}{1+\frac{k}{n}} + v_{\{1,\ldots,m\}}\prod_{k=1}^{m-1} \frac{1}{1+\frac{k}{n}}
\end{align}

The next test function of interest then is 
\begin{align} \label{eqn:X_N(1,ldots,m)unidirectionalring}
X_{N(\{1,\ldots,m\})}' = \begin{cases} 
0, & i=0,j\in S_k,\ell=j\\
X_{n}, & i=n,j=1,\ell=j\\
X_{N(\{1,\ldots,m\})}, & \text{otherwise}
\end{cases}
\end{align}
which combined with (\ref{eqn:hespanha_eqn}) and (\ref{eqn:lambda_ij_unidirectionalring}) gives
\begin{align}\label{eqn:v_1,ldots,m_unidirectionalring}
    v_{\{1,\ldots,m\}}&=\frac{1+p\lambda v_{n}}{p\lambda+\frac{m\lambda}{n}} 
\end{align}

Clearly
\begin{align}
     \frac{1+p\lambda v_{n}}{\lambda+\frac{m\lambda}{n}} \leq v_{\{1,\ldots,m\}} \leq \frac{1+p\lambda v_{n}}{p(\lambda+\frac{m\lambda}{n})}
\end{align}
which coupled with (\ref{eqn:v_m_obtained_recursively_unidirectionalring}) gives

\begin{align}\label{eqn:v_m_bounds_unidirectional_ring}
    \frac{1}{\lambda} \sum_{j=1}^{m} \prod_{k=1}^{j} \frac{1}{1+\frac{k}{n}} + pv_{n}\prod_{k=1}^{m} \frac{1}{1+\frac{k}{n}} \leq v_m \leq \frac{1}{p\lambda} \sum_{j=1}^{m} \prod_{k=1}^{j} \frac{1}{1+\frac{k}{n}} + v_{n}\prod_{k=1}^{m} \frac{1}{1+\frac{k}{n}}
\end{align}

The last useful test function is
\begin{align}
X_{n}' = \begin{cases} 
0, & i=0,j\in S_k,\ell=j\\
X_{n-1}, & i=n-1,j=n,\ell=j\\
X_{N(\{1,\ldots,m\})}, & \text{otherwise}
\end{cases}
\end{align}
which combined with (\ref{eqn:hespanha_eqn}) and (\ref{eqn:lambda_ij_unidirectionalring}) gives
\begin{align}\label{eqn:v_n_unidirectionalring}
    v_{n}&=\frac{1+(1-q)\lambda v_{n-1}}{(1-q)\lambda+\frac{\lambda}{n}} \nonumber \\
    &=\frac{\frac{1}{\lambda}}{1-q+\frac{1}{n}} +\frac{(1-q) v_{n-1}}{1-q+\frac{1}{n}}    
\end{align}

Before we move forward to analyze the adversarial effect on age scaling at regular nodes for different cases of probabilities $p$ and $q$, we state the following two lemmas that will be helpful in our derivations. Lemma~\ref{lemma:sumprod_upperbound_lowerbound} and Lemma~\ref{lemma:sumprod_upperbound} below are refined and generalized versions of \cite[Lemma~2]{baturalp21comm_struc}. Our results are more precise and general as they allow summations to go up to an arbitrary $n_0$ which is a function of $n$ and give order-wise exact expressions.

\begin{lemma}\label{lemma:sumprod_upperbound_lowerbound}
Let $n_0=\omega(\sqrt{n})$, then $\sum_{j=1}^{n_0}  \bigg( \prod_{k=1}^{j} \frac{1}{1+\frac{k}{n}} \bigg)$ is both $O(\sqrt{n})$ and $\Omega(\sqrt{n})$.
\end{lemma}

\begin{Proof}
Since
\begin{align}
    \frac{x}{1+x} \leq \log(1+x) \leq x
\end{align}
we choose $x=\frac{k}{n}$ to get
\begin{align}
    \frac{\frac{k}{n}}{1+\frac{k}{n}} \leq \log \bigg( 1+\frac{k}{n} \bigg) \leq \frac{k}{n} \nonumber \\
    \frac{k}{2n}  \leq \frac{k}{n+k} \leq \log \bigg( 1+\frac{k}{n} \bigg) \leq \frac{k}{n} 
\end{align}
Summing over $k$ gives
\begin{align}
    \frac{1}{2}\sum_{k=1}^{j} \frac{k}{n}  \leq \sum_{k=1}^{j} \log\bigg( 1+\frac{k}{n} \bigg) \leq \sum_{k=1}^{j} \frac{k}{n} 
\end{align}
Using $\sum_{k=1}^{j} \frac{k}{n} = \frac{j(j+1)}{2n}$ and $ \frac{j^2}{2n} \leq \frac{j(j+1)}{2n} \leq \frac{j^2}{n} $ gives the following bounds
\begin{align} \label{eqn:inequatlities_sumlog}
    \frac{j^2}{4n}  \leq \sum_{k=1}^{j} \log\bigg( 1+\frac{k}{n} \bigg) \leq \frac{j^2}{n}
\end{align}
Note that
\begin{align}
    -\log \bigg( \prod_{k=1}^{j} \frac{1}{1+\frac{k}{n}} \bigg)= \sum_{k=1}^{j} \log \bigg( 1+\frac{k}{n} \bigg)
\end{align}
and therefore, raising all sides to negative exponent in (\ref{eqn:inequatlities_sumlog}) gives
\begin{align}\label{eqn:inequalities_exponential_bounds}
    e^{-\frac{j^2}{n}} \leq  \prod_{k=1}^{j} \frac{1}{1+\frac{k}{n}}  \leq e^{-\frac{j^2}{4n}}
\end{align}
and consequently
\begin{align}\label{eqn:inequalities_1uponsqrtn_summation}
    \frac{1}{\sqrt{n}}\sum_{j=1}^{n_0}   e^{-\frac{j^2}{n}} \leq \frac{1}{\sqrt{n}}\sum_{j=1}^{n_0}  \bigg( \prod_{k=1}^{j} \frac{1}{1+\frac{k}{n}} \bigg) \leq \frac{1}{\sqrt{n}}\sum_{j=1}^{n_0}   e^{-\frac{j^2}{4n}} 
\end{align}
Finally, note that, for any constant $C$, using upper and lower Riemann sums, we get
\begin{align}
    \int_{\frac{1}{\sqrt{n}}}^{\frac{n_0}{\sqrt{n}}}e^{-\frac{t^2}{C}}dt \leq \frac{1}{\sqrt{n}}\sum_{j=1}^{n_0} e^{-\frac{j^2}{Cn}} \leq \int_{0}^{\frac{n_0}{\sqrt{n}}}e^{-\frac{t^2}{C}}dt
\end{align}
Hence, for $n_0=\omega(\sqrt{n})$ and for large $n$,
\begin{align}
     \frac{1}{\sqrt{n}}\sum_{j=1}^{n_0}  e^{-\frac{j^2}{Cn}} = \int_{0}^{\infty}e^{-\frac{t^2}{C}}dt = \frac{\sqrt{C\pi}}{2}
\end{align}
Thus, (\ref{eqn:inequalities_1uponsqrtn_summation}) results in $\sum_{j=1}^{n_0}  \bigg( \prod_{k=1}^{j} \frac{1}{1+\frac{k}{n}} \bigg)$ being both $O(\sqrt{n})$ and $\Omega(\sqrt{n})$.
\end{Proof}

\begin{lemma}\label{lemma:sumprod_upperbound}
Let $n_0=O(\sqrt{n})$, then $\sum_{j=1}^{n_0}  \bigg( \prod_{k=1}^{j} \frac{1}{1+\frac{k}{n}} \bigg)$ is $O(\sqrt{n})$.
\end{lemma}
The proof of Lemma~\ref{lemma:sumprod_upperbound} follows similarly to the proof of Lemma~\ref{lemma:sumprod_upperbound_lowerbound}.

We now analyze how the choice of probabilities $p$ and $q$ by the adversary in unidirectional ring affect the expected age at regular nodes in the regime of large $n$.

\begin{figure}[t]
 	\begin{center}
 	\subfigure[]{\includegraphics[width=.42\textwidth]{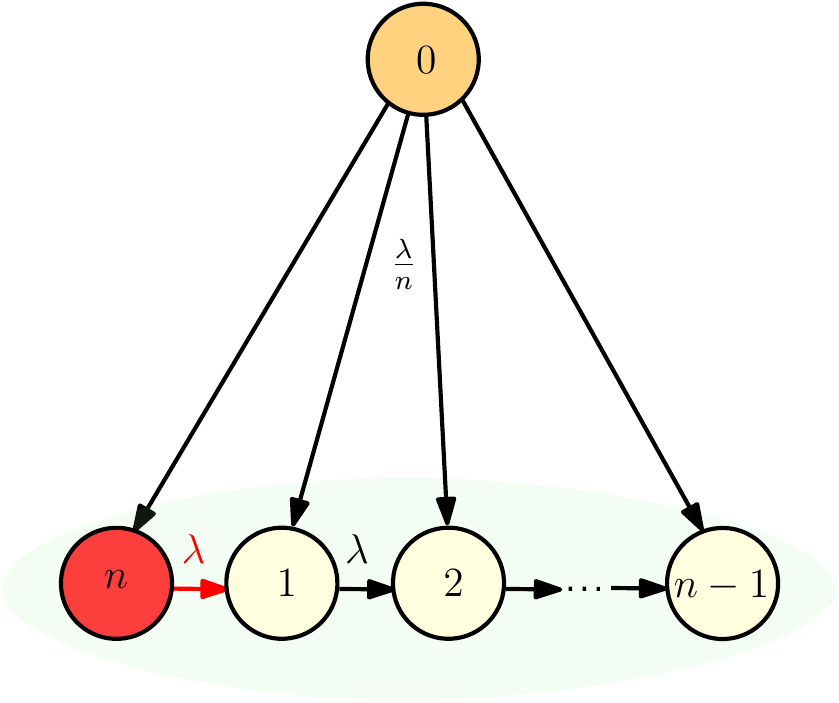}}
        \qquad \qquad \quad
 	\subfigure[]{\includegraphics[width=.42\textwidth]{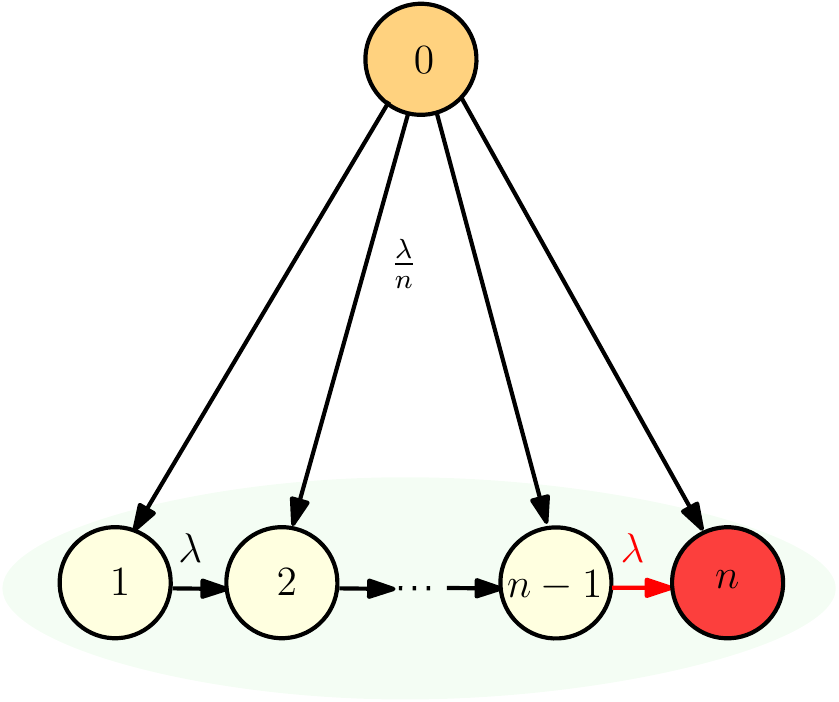}}
 	\end{center}
 	\caption{Line network transformations of unidirectional ring.}
 	\label{fig:unidirectional_ring_cutlines}
\end{figure}

\subsection{URN Case~1: $p>0,q=1$}\label{subsec:uniring_case1}

This corresponds to the case where the infected node $n$ does not receive anything from the gossip process but sends out packets, with timestamps manipulated to current time $t$, to node $1$ according Poisson process with rate $p\lambda$. Since the network topology under consideration is a unidirectional ring, this case essentially converts the ring network of Fig.~\ref{fig:unidirectional_ring_model} into the line network model shown in Fig.~\ref{fig:unidirectional_ring_cutlines}(a). In this case, (\ref{eqn:v_n_unidirectionalring}) gives $v_n=\frac{n}{\lambda}=O(n)$, since the infected node is an isolated node that only receives updates from the source. Further, from (\ref{eqn:v_m_bounds_unidirectional_ring}) we have
\begin{align}\label{eqn:v_m_bounds_case1_unidirectional_ring}
    \frac{1}{\lambda} \sum_{j=1}^{m} \prod_{k=1}^{j} \frac{1}{1+\frac{k}{n}} + p\frac{n}{\lambda}\prod_{k=1}^{m} \frac{1}{1+\frac{k}{n}} \leq v_m \leq \frac{1}{p\lambda} \sum_{j=1}^{m} \prod_{k=1}^{j} \frac{1}{1+\frac{k}{n}} + \frac{n}{\lambda}\prod_{k=1}^{m} \frac{1}{1+\frac{k}{n}}
\end{align}

The presence of adversary at node $n$ breaks the symmetry in the ring and the age processes at all the regular nodes are no longer statistically similar.
Therefore we analyze how expected age $v_m$ scales at node index $m$, where $m$ scales as a specific function of network size $n$. 

\emph{URN Case~1(a): $m= O(n^{\frac{1}{2}})$.} For the first term in the bounds of (\ref{eqn:v_m_bounds_case1_unidirectional_ring}), we know that $\sum_{j=1}^{m} \prod_{k=1}^{j} \frac{1}{1+\frac{k}{n}}$ is $O(\sqrt{n})$ from Lemma~\ref{lemma:sumprod_upperbound}. The second term in (\ref{eqn:v_m_bounds_case1_unidirectional_ring}) has product of $n$ and $\prod_{k=1}^{m} \frac{1}{1+\frac{k}{n}}$. From (\ref{eqn:inequalities_exponential_bounds}), we know $e^{-\frac{m^2}{n}} \leq  \prod_{k=1}^{m} \frac{1}{1+\frac{k}{n}}  \leq e^{-\frac{m^2}{4n}}$. Since $m= O(\sqrt{n})$, for any constant $C$, $e^{-\frac{m^2}{Cn}}$ converges to a non-zero constant for large $n$, making the second term $O(n)$. Hence, we conclude from the second term of (\ref{eqn:v_m_bounds_case1_unidirectional_ring}) that $v_m$ is both $O(n)$ and $\Omega(n)$, and expected age at node $m$ scales as $n$.

\emph{URN Case~1(b): $m =\Omega(n^{\frac{1}{2}+ \epsilon})$, $\epsilon>0$.} In this case, Lemma~\ref{lemma:sumprod_upperbound_lowerbound} provides that $\sum_{j=1}^{m} \prod_{k=1}^{j} \frac{1}{1+\frac{k}{n}}$ is both $O(\sqrt{n})$ and $\Omega(\sqrt{n})$. Further, for any constant $C$, both $e^{-\frac{m^2}{Cn}}$ and $n e^{-\frac{m^2}{Cn}}$ converge to $0$ as $n$ becomes large. Hence, the first term of (\ref{eqn:v_m_bounds_case1_unidirectional_ring}) provides that expected age at node $m$ scales as $\sqrt{n}$, and thus, $v_m$ is both $O(\sqrt{n})$ and $\Omega(\sqrt{n})$.

\emph{URN Case~1(c): $m =\Omega(n^{\frac{1}{2}+ \epsilon (n)})$, $\epsilon (n)=\frac{\log \alpha+\log \log n}{2\log n}$, $\alpha > 0$.} This case gives a flavor of the transitional behavior between \emph{Case~1(a)} and \emph{Case~1(b)} above. From (\ref{eqn:inequalities_exponential_bounds}), $\prod_{k=1}^{j} \frac{1}{1+\frac{k}{n}}$ scales as $  e^{-\frac{j^2}{\beta n}}$, $\beta \in [1,4]$, and consequently, the expression $n\prod_{k=1}^{m} \frac{1}{1+\frac{k}{n}}$ in the second term in the upper bound and lower bound of (\ref{eqn:v_m_bounds_case1_unidirectional_ring}) becomes
\begin{align}\label{eqn:uniring_case1_part1}
    n\prod_{k=1}^{m} \frac{1}{1+\frac{k}{n}} \approx  ne^{-\frac{m^2}{\beta n}} = ne^{-\frac{\alpha n \log n}{\beta n}} = nn^{-\frac{\alpha}{\beta} } =n^{1-\frac{\alpha}{\beta}} = n^{1-\gamma} 
\end{align}
When $1-\gamma >\frac{1}{2}$, which is true for smaller values of $\alpha$, then the expected age $v_m$ at node index $m$ scales as $n^{1-\gamma} $, since the first term in the bounds of (\ref{eqn:v_m_bounds_case1_unidirectional_ring}) is  at most $O(\sqrt{n})$.

\subsection{URN  Case~2: $p=0, q \leq 1$} \label{subsec:uniring_case2}

This corresponds to the case where no timestomped packets are effectively injected into the gossiping process, since all packets sent by the infected node $n$ carry a timestamp of $0$ causing them get rejected at the receiving node $1$, even when the received packet has fresh content. This effectively converts the ring network of Fig.~\ref{fig:unidirectional_ring_model} into the line network of Fig.~\ref{fig:unidirectional_ring_cutlines}(b). We now analyze how absence of a contributing node before node $1$ affects scaling behaviour of the expected age at other regular nodes of the network. In this case, (\ref{eqn:v_1,ldots,m_unidirectionalring}) becomes $v_{\{1,\ldots,m\}}=\frac{n}{m\lambda}$ which transforms (\ref{eqn:v_m_obtained_recursively_unidirectionalring}) into
\begin{align} \label{eqn:v_m_noadversarycontri_unidirectionalring}
    v_m= \frac{1}{\lambda} \left( \sum_{j=1}^{m-1} \prod_{k=1}^{j} \frac{1}{1+\frac{k}{n}} + \frac{n}{m}\prod_{k=1}^{m-1} \frac{1}{1+\frac{k}{n}} \right )
\end{align}
We now look at expected age $v_m$ when $m$ scales as a specific function of network size $n$. 

\emph{URN Case~2(a): $m=o(\sqrt{n})$.} The first term in (\ref{eqn:v_m_noadversarycontri_unidirectionalring}) is $O(\sqrt{n})$ from Lemma~\ref{lemma:sumprod_upperbound}. The second term can be bounded from above and below by expressions of form $\frac{n}{m} e^{-\frac{(m-1)^2}{Cn}}$ due to (\ref{eqn:inequalities_exponential_bounds}), where $C$ is some constant. In this case, in the regime of large $n$, $e^{-\frac{(m-1)^2}{Cn}}$ converges to $e^0=1$, and hence, the age scaling in this case is determined by the fraction $\frac{n}{m}$ and is at least $\omega(\sqrt{n})$. For example, if $m= n^{\alpha}, \alpha \in [0,\frac{1}{2})$, then $v_m$ scales as $O(n^{1-\alpha})$ and $\Omega(n^{1-\alpha})$.

\emph{URN Case~2(b): $m=\Omega(\sqrt{n})$.} Here in (\ref{eqn:v_m_noadversarycontri_unidirectionalring}), Lemma~\ref{lemma:sumprod_upperbound_lowerbound} provides that $\sum_{j=1}^{m-1} \prod_{k=1}^{j} \frac{1}{1+\frac{k}{n}}$ is $O(\sqrt{n})$. Further $\frac{n}{m} e^{-\frac{(m-1)^2}{Cn}}$ is $O(\sqrt{n})$, since $e^{-\frac{(m-1)^2}{Cn}}$ converges to a constant and $\frac{n}{m}$ is $O(\sqrt{n})$.

\subsection{URN Case~3: $p>0,q<1$} \label{subsec:uniring_case3}

This corresponds to the case where the infected node both partially receives packets from node $n-1$ and partially sends packets to node $1$ carrying timestamps manipulated to $t$. This case is particularly interesting, for it highlights how the adversary fails to deteriorate age scaling at regular nodes when it both sends and receives packets via gossiping at the same time.

We see from (\ref{eqn:v_n_unidirectionalring}) that $v_n$ is dependent on $v_{n-1}$, which in turn can be obtained using a combination of (\ref{eqn:v_m_obtained_recursively_unidirectionalring}) and (\ref{eqn:v_1,ldots,m_unidirectionalring}) as follows
\begin{align}
    v_{n-1}=& \frac{1}{\lambda} \sum_{j=1}^{n-2} \prod_{k=1}^{j} \frac{1}{1+\frac{k}{n}} + v_{\{1,\ldots,n-1\}}\prod_{k=1}^{n-2} \frac{1}{1+\frac{k}{n}} \nonumber \\
    =& \frac{1}{\lambda} \sum_{j=1}^{n-2} \prod_{k=1}^{j} \frac{1}{1+\frac{k}{n}} + \frac{1+p\lambda v_{n}}{p\lambda+\frac{(n-1)\lambda}{n}} \prod_{k=1}^{n-2} \frac{1}{1+\frac{k}{n}} 
\end{align}
thereby giving $v_{n-1}$ in terms of $v_n$, which can be plugged back into (\ref{eqn:v_n_unidirectionalring}) as
\begin{align}
    v_{n}&= \frac{\frac{1}{\lambda}}{1-q+\frac{1}{n}} +\frac{(1-q) }{1-q+\frac{1}{n}}\left( \frac{1}{\lambda} \sum_{j=1}^{n-2} \prod_{k=1}^{j} \frac{1}{1+\frac{k}{n}} + \frac{1+p\lambda v_{n}}{p\lambda+\frac{(n-1)\lambda}{n}} \prod_{k=1}^{n-2} \frac{1}{1+\frac{k}{n}}  \right)
\end{align}
to obtain age of $v_n$ as
\begin{align}
    v_n = \frac{1}{\lambda}. \frac{\frac{1}{1-q+\frac{1}{n}} + \frac{(1-q) }{1-q+\frac{1}{n}}\left(  \sum_{j=1}^{n-2} \prod_{k=1}^{j} \frac{1}{1+\frac{k}{n}} + \frac{1}{p+\frac{(n-1)}{n}} \prod_{k=1}^{n-2} \frac{1}{1+\frac{k}{n}}  \right)}{1-\frac{(1-q) }{(1-q)+\frac{1}{n}} \left( \frac{p }{p+\frac{(n-1)}{n}} \prod_{k=1}^{n-2} \frac{1}{1+\frac{k}{n}}  \right)}
\end{align}
Using $\frac{(1-q) }{(1-q)+\frac{1}{n}} \leq 1$ and $\prod_{k=1}^{n-2} \frac{1}{1+\frac{k}{n}}  \leq 1$, we obtain the following upper bound
\begin{align}
    v_n \leq \frac{1}{\lambda}. \frac{\frac{1}{1-q+\frac{1}{n}} + \left(  \sum_{j=1}^{n-2} \prod_{k=1}^{j} \frac{1}{1+\frac{k}{n}} + \frac{1}{p+\frac{(n-1)}{n}}   \right)}{1-\left( \frac{p }{p+\frac{(n-1)}{n}} \right)}
\end{align}

We first find the age scaling at the infected node $v_n$ in the regime of large network size $n$. The bounds $\frac{n-1}{n} \geq \frac{1}{2}$ for $n \geq 2$, $\frac{1}{1-q+\frac{1}{n}}\leq \frac{1}{1-q}$, and $\sum_{j=1}^{n-2} \prod_{k=1}^{j} \frac{1}{1+\frac{k}{n}} \leq \sum_{j=1}^{n} \prod_{k=1}^{j} \frac{1}{1+\frac{k}{n}}$ provide the following further simplification of the upper bound
\begin{align} \label{eqn:v_n_bounded_both_input_output_unidirectional_ring}
    v_n \leq \frac{1}{\lambda}. \frac{\frac{1}{1-q} + \left(  \sum_{j=1}^{n} \prod_{k=1}^{j} \frac{1}{1+\frac{k}{n}} + \frac{1}{p+\frac{1}{2}}   \right)}{1-\left( \frac{p }{p+\frac{1}{2}} \right)}
\end{align}

Since $q <1 $ and $p>0$ in this case, the above mentioned bound comprises primarily constants and age is thus determined by the term $\sum_{j=1}^{n} \prod_{k=1}^{j} \frac{1}{1+\frac{k}{n}}$, which is $O(\sqrt{n})$ from Lemma~\ref{lemma:sumprod_upperbound_lowerbound}. Hence, irrespective of how large $q$ is, the age at the infected node still scales as $O(\sqrt{n})$ instead of $O(n)$. This observation is similar to what was seen in fully connected networks, where we observed that when the infected node accepts packets via gossiping, then the gossip network brings down the age of the infected node from $O(n)$ to $O(\log n)$ characteristic of fully connected networks, the analogue of which is $O(\sqrt{n})$ for the unidirectional ring.

Now, we proceed to see how age scales at a typical regular node on the ring, for which we consider individual terms in the upper and lower bound of (\ref{eqn:v_m_bounds_unidirectional_ring}). The first term comprises $\sum_{j=1}^{m} \prod_{k=1}^{j} \frac{1}{1+\frac{k}{n}}$ which is $O(\sqrt{n})$ irrespective of node index $m$ from Lemmas~\ref{lemma:sumprod_upperbound_lowerbound}~and~\ref{lemma:sumprod_upperbound}. The second term contains a product of $v_{n}$ and $\prod_{k=1}^{m} \frac{1}{1+\frac{k}{n}}$, where the former is $O(\sqrt{n})$ from (\ref{eqn:v_n_bounded_both_input_output_unidirectional_ring}) and the later is bounded from (\ref{eqn:inequalities_exponential_bounds}) by $e^{-\frac{m^2}{4n}}\leq 1$. Hence, $v_m$ is $O(\sqrt{n})$ for each node index $m$.

\subsection{Key Insights}  

\emph{URN Case~2} highlights an interesting attribute of the expected age in the unidirectional ring. Upon cutting one node-to-node link on the ring, while the age still scales as $O(n)$ at the first exposed node, it transitions (more gradually in \emph{URN Case~2} than \emph{URN Case~1}) to $O(\sqrt{n})$ until $m=o(\sqrt{n})$, after which it stays at $O(\sqrt{n})$ for the rest of the ring. Note that in the absence of any adversary or jammed link, \cite{baturalp21comm_struc} shows that age scales as $\sqrt{n}$ at all nodes of a unidirectional ring. Though node $1$ in \emph{URN Case~2} could not receive any packets from the gossip process and had to rely on occasional updates from the source, the missing link before node $1$ affects the age in only the initial portion of the ring and for nodes of higher indices the age still scales as $O(\sqrt{n})$. Further, we note that $v_{n}$ is $O(\sqrt{n})$ for $q<1$ and $O(n)$ for $q=1$ in \emph{URN Case~2}, this can be obtained by noting that $v_{n-1}=O(\sqrt{n})$ from \emph{URN Case~2(b)} since $n-1=\Omega(\sqrt{n})$, and using $\frac{1}{n} \geq 0$ in (\ref{eqn:v_n_unidirectionalring}).

Finally, its worthwhile to note that the maximum age deterioration in the unidirectional ring is caused when the adversary chooses to actively send out packets with timestamp $t$ to the following node $1$ and reject all incoming packets from preceding node $n-1$, i.e., $q=1$ and $p=1$, a special case of \emph{URN Case~1}. This is because $v_n=\frac{n}{\lambda}= O(n)$ and for this case $v_{\{1,\ldots,m\}}$ in  (\ref{eqn:v_1,ldots,m_unidirectionalring}) (and consequently $v_m$ in (\ref{eqn:v_m_obtained_recursively_unidirectionalring})) turns out to be an increasing function of $p$, a fact that can be verified by taking first derivative of $v_{\{1,\ldots,m\}}$ in  (\ref{eqn:v_1,ldots,m_unidirectionalring}) with respect to $p$, which justifies choice of $p=1$. Further, in \emph{URN Case~1}, which corresponds to $q=1$, the expected age $v_m$ is $O(n)$ for significantly larger portion of the ring than in \emph{URN Case~2}, where the expected age $v_m$ reduces form $O(n)$ to $O(\sqrt{n})$ more gradually. Finally, we note that for any meaningful age scaling deterioration, the adversary must choose between \emph{URN Case~1} and \emph{URN Case~2}, i.e., the infected node should either not accept packets at all but only inject packets in the gossiping process, or not inject any packets at all and thereby limit its contribution completely. A mix of these two actions is not optimal, as seen in \emph{URN Case~3}, where age scales as $O(\sqrt{n})$ at all nodes.\footnote{Another interesting network to study is the bidirectional ring network, where each node transmits packets to two other nodes with rate $\frac{\lambda}{2}$. In the absence of any adversarial actions, it is known that both unidirectional and bidirectional ring networks have exactly the same age for all $n$ \cite{baturalp21comm_struc}. Though it would be interesting to see in the presence of adversarial actions, if the infected node still fails to propagate its effect on age scaling of a large portion of the network, the manipulated timestamps and lack of symmetry in the network due to the presence of an adversary make it particularly difficult to analyze the bidirectional ring, and we leave it as a future direction for now.}

\section{Numerical Results}

\begin{figure}[t]
\centerline{\includegraphics[width=0.6\linewidth]{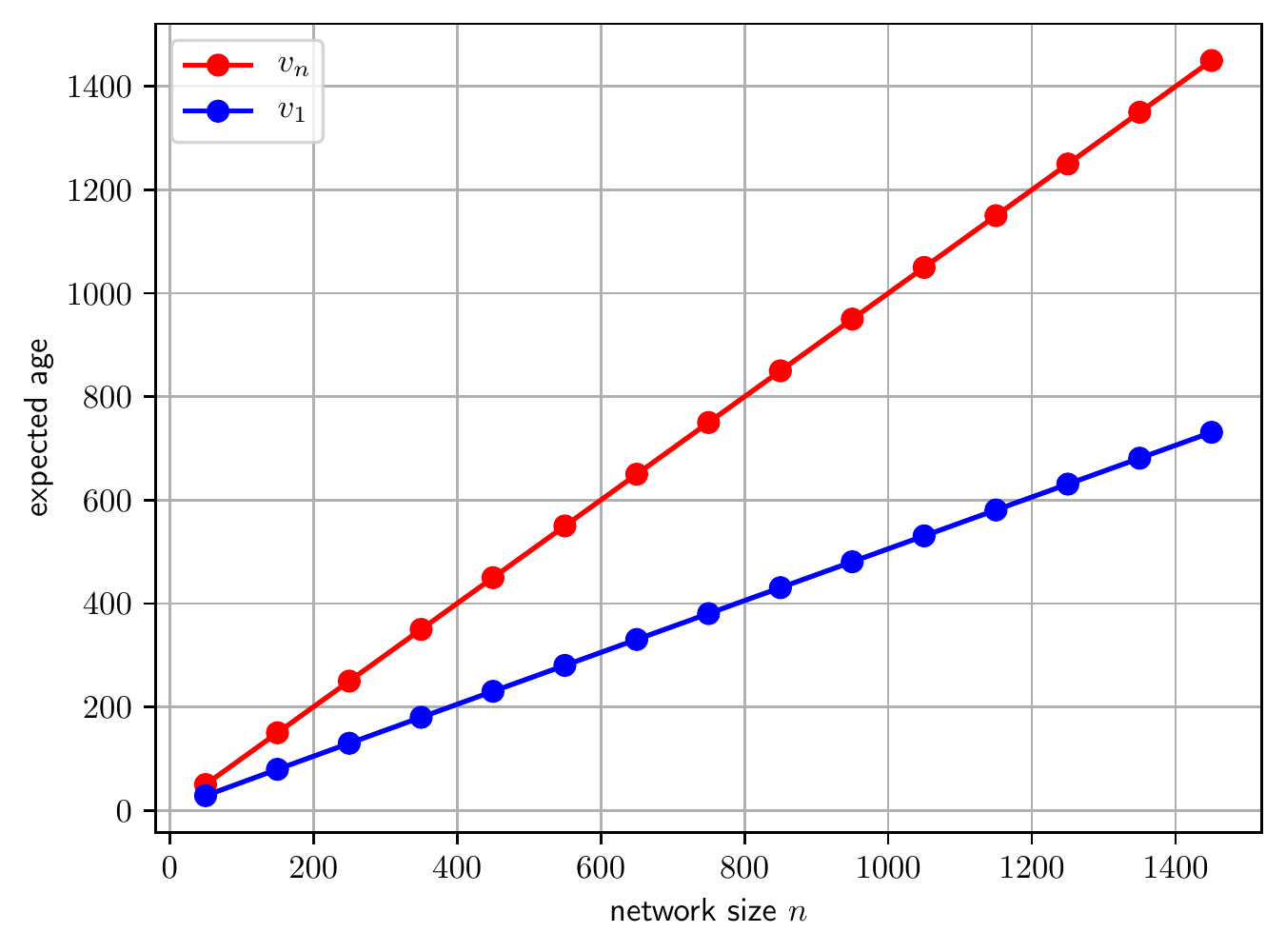}}
\caption{Node capture attack on fully connected network with $p=0.5$ and $q=1$.}
\label{fig:graph_case1_fullyconnec_q1_p05}
\end{figure}

\begin{figure}[t]
\centerline{\includegraphics[width=0.6\linewidth]{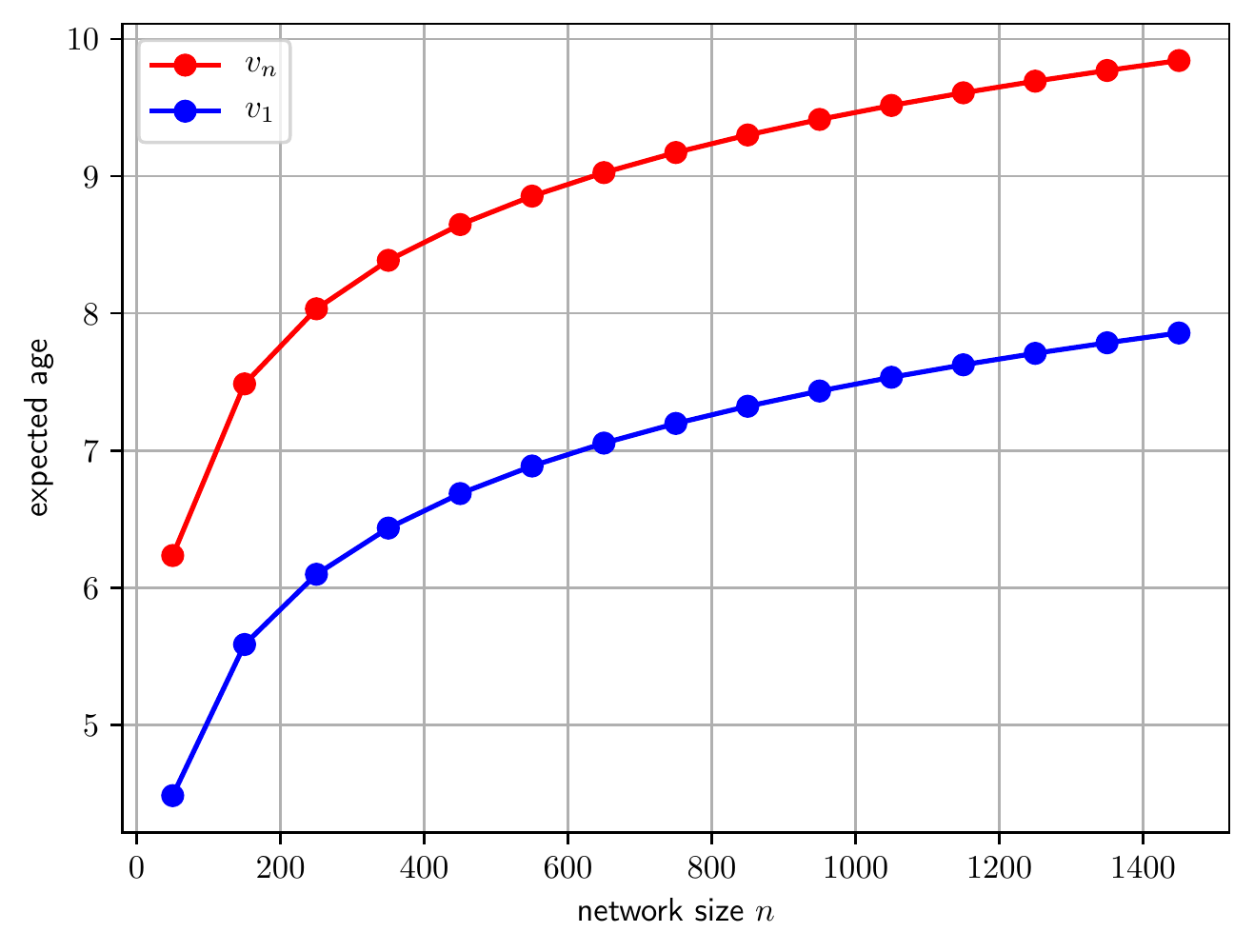}}
\caption{Node capture attack on fully connected network with $p=0$ and $q=0.5$.}
\label{fig:graph_case1_fullyconn_q1_p05}
\vspace*{-0.4cm}
\end{figure}

\begin{figure}[t]
\centerline{\includegraphics[width=0.6\linewidth]{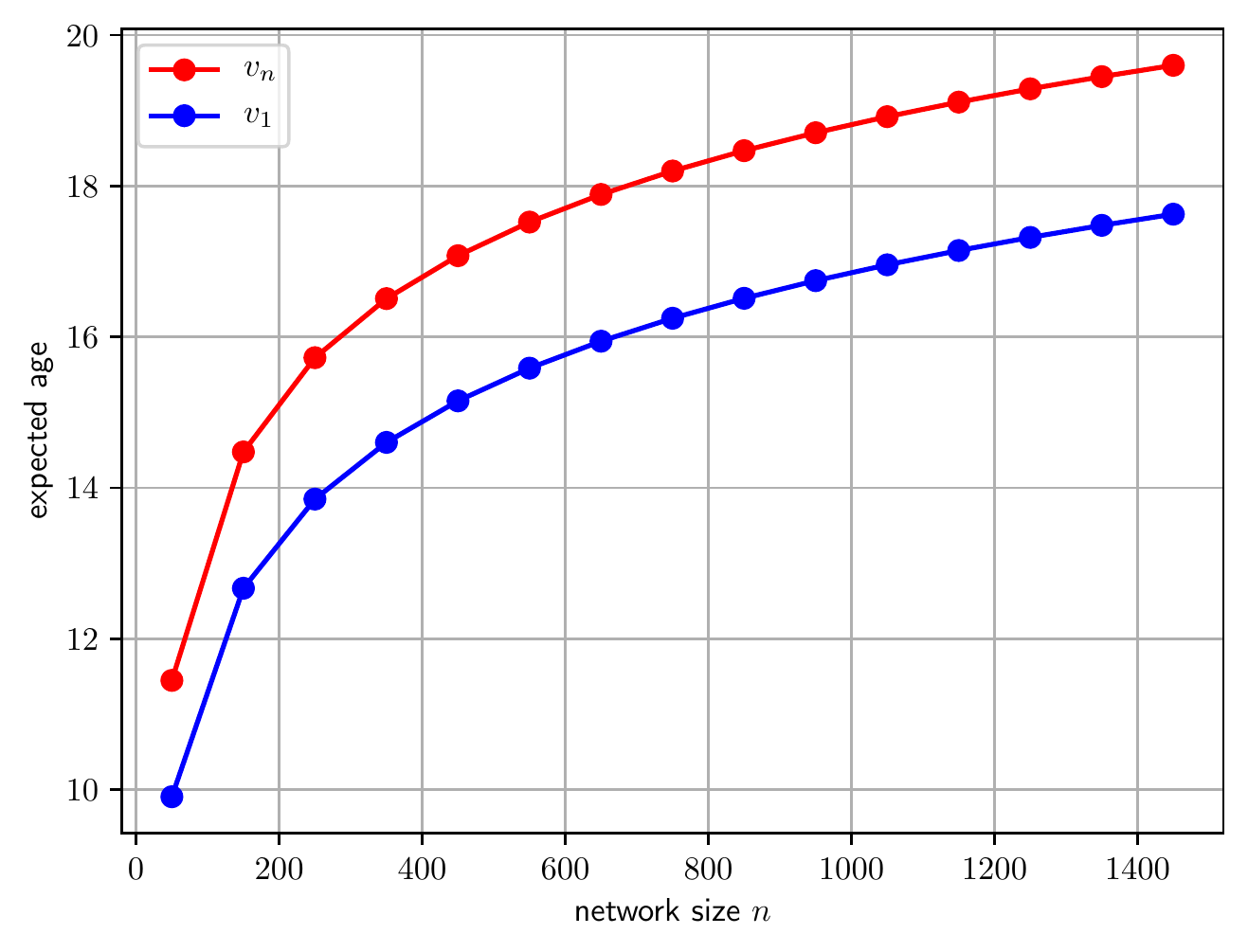}}
\caption{Node capture attack on fully connected network with $p=0.5$ and $q=0.5$.}
\label{fig:graph_case1_fullyconn_q05_p05}
\end{figure}

\begin{figure}[t]
\centerline{\includegraphics[width=0.6\linewidth]{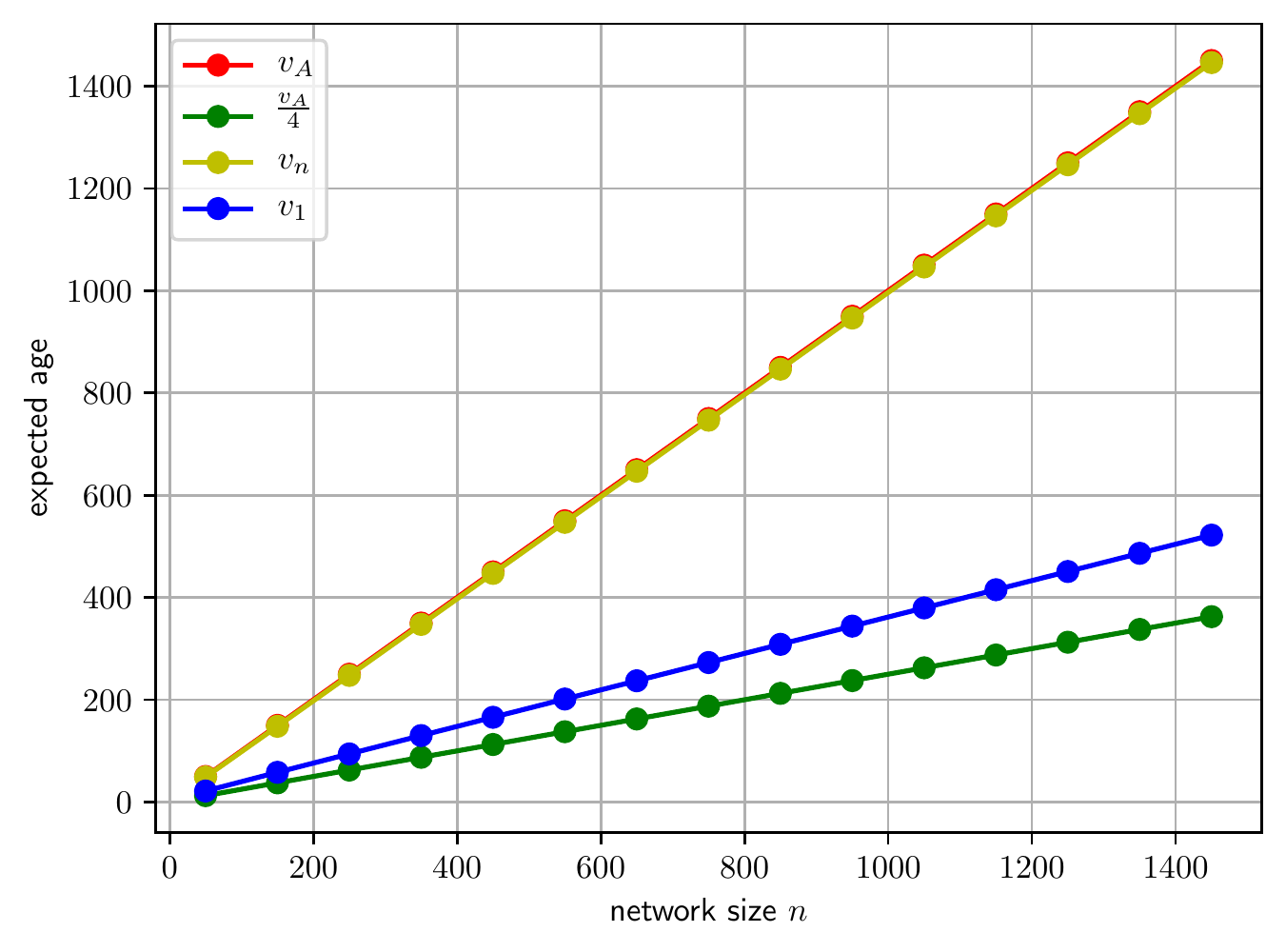}}
\caption{MITM attack on fully connected network of $n$ nodes.}
\label{fig:graph_fully_conn_net_MITM_attack}
\vspace*{-0.4cm}
\end{figure}

\begin{figure}[t]
\centerline{\includegraphics[width=0.6\linewidth]{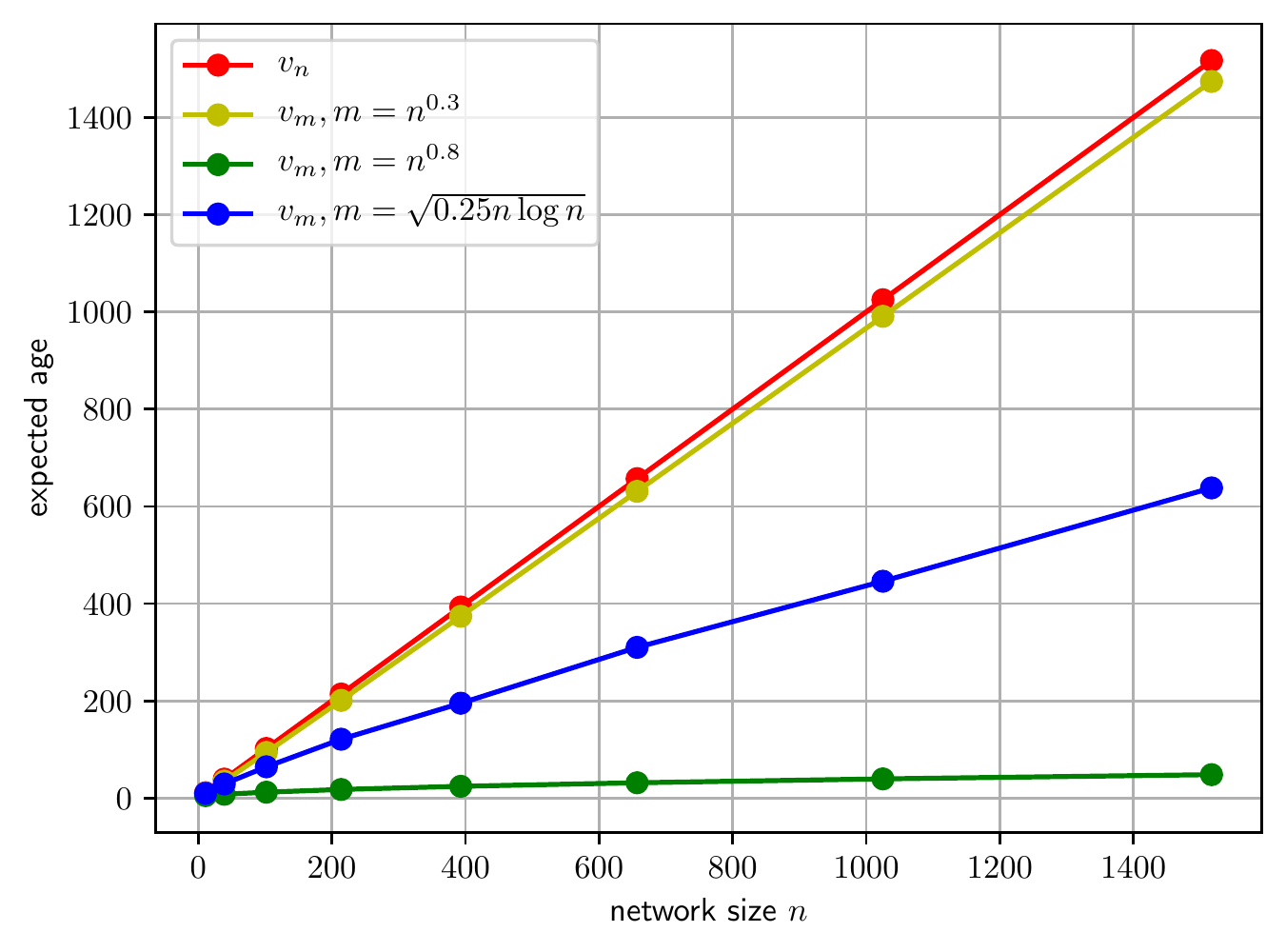}}
\caption{Node capture attack on unidirectional ring with $p=0.5$ and $q=1$.}
\label{fig:graph_case1_unidrectional_ring_q1_p05}
\end{figure}

\begin{figure}[t]
\centerline{\includegraphics[width=0.6\linewidth]{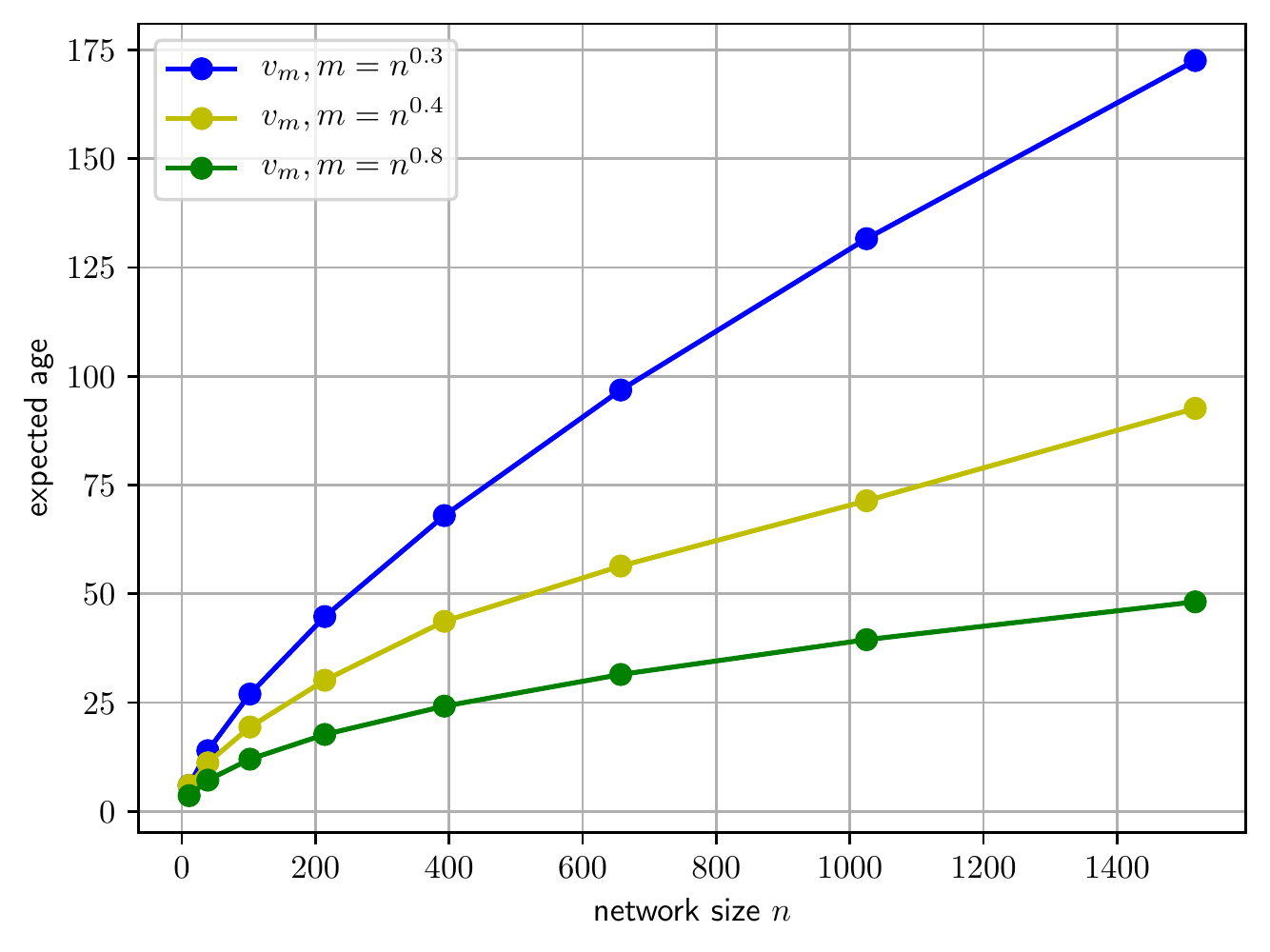}}
\caption{Node capture attack on unidirectional ring with $p=0$ and $q=0.5$.}
\label{fig:graph_case2_unidrectional_ring_q1_p05}
\vspace*{-0.4cm}
\end{figure}

\begin{figure}[t]
\centerline{\includegraphics[width=0.6\linewidth]{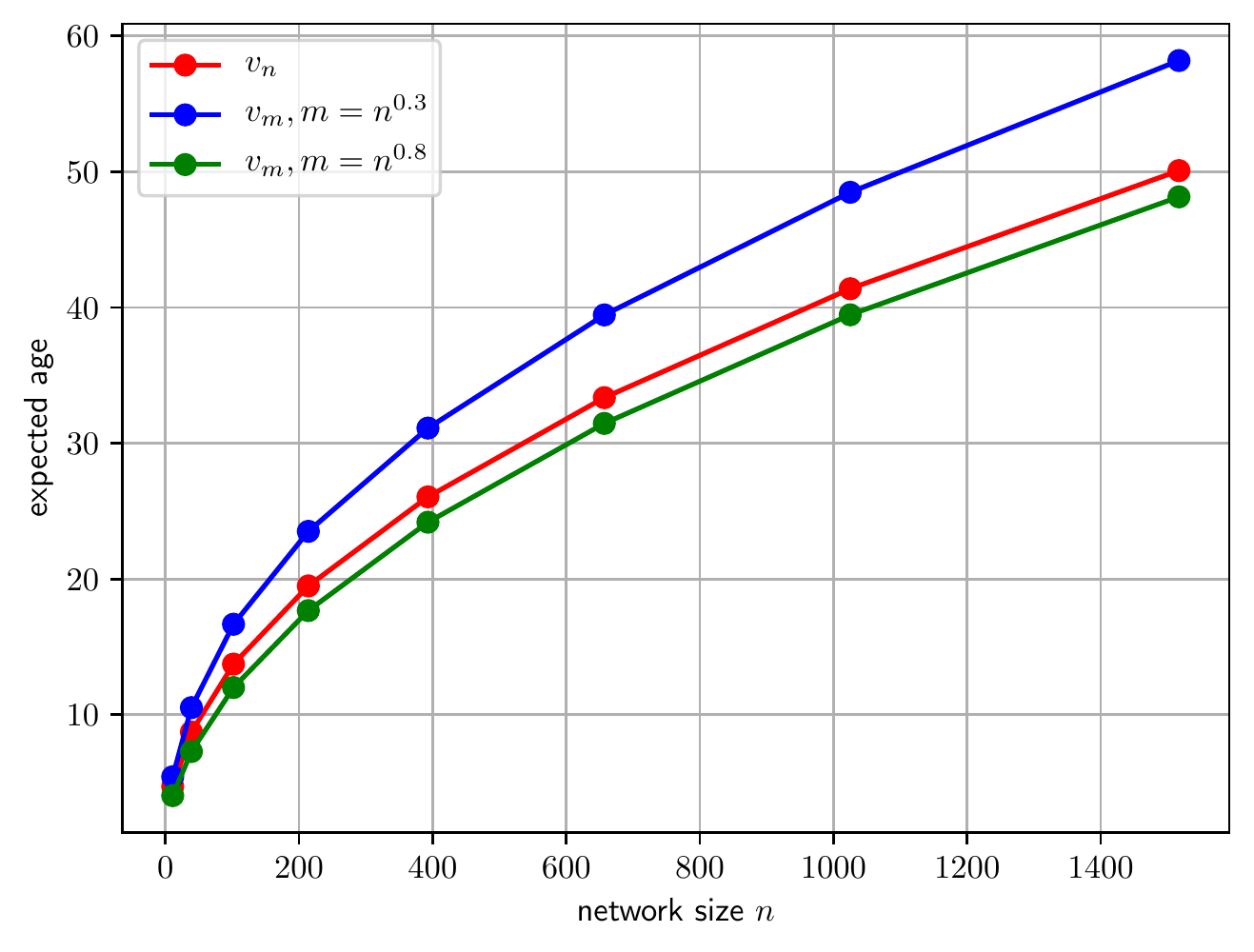}}
\caption{Node capture attack on unidirectional ring with $p=0.5$ and $q=0.5$.}
\label{fig:graph_case3_unidrectional_ring_q1_p05}
\end{figure}

We first provide numerical results for all three cases of Section~\ref{sect:no-capture} corresponding to fully connected networks (FCNs), choosing $\lambda=1$ for the purposes of this section. Fig.~\ref{fig:graph_case1_fullyconnec_q1_p05} shows the expected age at a regular node $v_1$ and the infected node $v_n$ in \emph{FCN Case~1}, employing choice of probabilities $p=0.5$ and $q=1$. The age at the infected node $v_n$ in red color grows as $\frac{n}{\lambda}$ as expected from (\ref{eqn:Xn_eclipse_adversary}) by plugging $q=1$. The age at a regular node $v_1$ in blue color grows linearly as $O(n)$ and lies broadly between
$pv_n$ and $\frac{pv_n}{2}$, as suggested by (\ref{eqn:case1_v_1_lowerbound_fullyconn}) and (\ref{eqn:case1_v_1_finalupperbound_fullyconn}).

Fig.~\ref{fig:graph_case1_fullyconn_q1_p05} shows the expected ages $v_1$ and $v_n$ for \emph{FCN Case~2}, using $p=0$ and $q=0.5$. The red and blue curves are roughly separated by a distance of about $\frac{1}{\lambda(1-q)}\approx 2$, given our choice of $\lambda=1$ and $q=0.5$, as suggested in (\ref{eqn:eclipse_vn_case2}), and both scale as logarithmically, with $v_1 \leq e \log n$ as suggested in (\ref{eqn:case2_v_1_upperbound_fullyconc}).

Finally, we plot $v_1$ and $v_n$ for \emph{FCN Case~3} in Fig.~\ref{fig:graph_case1_fullyconn_q05_p05}, choosing $p=0.5$ and $q=0.5$. Again the red and blue curves are roughly separated by a distance of $\frac{1}{\lambda(1-q)}\approx 2$ as suggested in (\ref{eqn:eclipse_vn_case2}). Both $v_1$ and $v_n$ scale as $O(\log n)$, with $v_1 \approx 2(\log n +1)$ as suggested in (\ref{eqn:case3_v_1_upperbound_fullyconc}).

Fig.~\ref{fig:graph_fully_conn_net_MITM_attack} shows the expected age at different nodes for Section~\ref{sect:mitmattack} when the adversary is positioned between the source and a node in a fully connected network, i.e., MITM attack. The green line shows the lower bound $\frac{v_A}{4}$ of (\ref{eqn:mitm_v1_lb_va_by_4}), where the age at a regular node $v_1$ lies above this lower bound. Adversary age $v_A$  grows as $\frac{n}{\lambda}=O(n)$ by virtue of being an isolated node. Finally, though (\ref{eqn:nmitm_vskn_lb_vA_by_2}) yields a loose lower bound of $\frac{v_A}{2}$, the graph shows that the age at the node that is in contact with the adversary, $v_n$, closely follows adversary age $v_A$.

Next, we present the set of numerical results for all three cases of Section~\ref{sect:unidirectionalring} corresponding to unidirectional ring networks (URNs). Fig.~\ref{fig:graph_case1_unidrectional_ring_q1_p05} demonstrates how expected age $v_m$ scales with network size $n$ for various index functions discussed in \emph{URN Case~1}. Notice that the data points are concentrated more towards the left side of the graph. This is because we choose node index $m=\lfloor n^{0.3} \rfloor$ casusing $m$ to change value only when $\lfloor n^{0.3} \rfloor$ changes its integer value, motivating us to choose $n=\lceil \ell^{\frac{1}{0.3}} \rceil$, $\ell \in \mathbb{N}$ to have $m$ increase consistently with $n$. 

Clearly, the $v_n$ in red color grows as $\frac{n}{\lambda}$ as stated in \emph{URN Case~1}. Next, we see that when $m=n^{0.3}=O(n^{\frac{1}{2}})$, the expected age $v_m$ in yellow color scales almost as $n$ since $e^{-\frac{m^2}{Cn}}$ converges to $0$ for larger values of $n$, which is in accordance with \emph{URN Case~1(a)}. \emph{URN Case~1(b)} is represented by the green curve, where $m=n^{0.8}=\Omega(n^{\frac{1}{2}+\epsilon})$ for $\forall \epsilon \in (0,0.3]$, and the corresponding $v_m$ scales as $1.25 \sqrt{n}$ similar to a unidirectional ring network without adversary of \cite{baturalp21comm_struc}. The blue curve shows \emph{URN Case~1} as a transition case between \emph{URN Case~1(a)} and \emph{URN Case~1(b)}. For choice of parameters $p=0.5$, $q=0.5$ and $\alpha=0.25$, $v_m$ is dictated primarily by the second term of (\ref{eqn:v_m_bounds_case1_unidirectional_ring}) to be $n^{0.875}$ with $\beta \approx 2$ as supported by (\ref{eqn:uniring_case1_part1}).

Fig.~\ref{fig:graph_case2_unidrectional_ring_q1_p05} demonstrates how expected age $v_m$ scales for various index functions discussed in \emph{URN Case~2}. The blue curve represents \emph{URN Case~2(a)} for choice of $\alpha=0.3$, and the term $n^{0.7}$ primarily dictates expected age $v_m$ for $m=n^{0.3}$. Likewise the yellow curve represents \emph{URN Case~2(a)} for choice of $\alpha=0.4$, causing $v_m$ to dominantly follow $n^{0.6}$, which illustrates how the age scaling behaviour gradually transitions across node indices in \emph{URN Case~2}. Lastly, the green curve exemplifies how for $n^{0.8}=\Omega(\sqrt{n})$, $v_m$ is practically unaffected by the presence of the adversary in \emph{URN Case~2(b)} and $v_m$ scales as approximately $1.25\sqrt{n}$ similar to a unidirectional ring network without adversary in \cite{baturalp21comm_struc}.

Lastly, for \emph{URN Case~3}, we observe in Fig.~\ref{fig:graph_case3_unidrectional_ring_q1_p05}  that the expected age scales as $O(\sqrt{n})$ at all nodes of the network. In accordance with (\ref{eqn:v_n_unidirectionalring}), the red curve $v_n$ closely follows the green curve $v_m$ for $m=n^{0.8}$, which almost scales as $1.25\sqrt{n}$  since all nodes of higher indices are almost unaffected by the adversary. Since packets are sent in a fixed direction around the ring, the blue curve representing $v_m$ for $m=n^{0.3}$ has higher expected age than $v_n$ at the infected node, owing to the presence of extra first term in (\ref{eqn:v_m_bounds_unidirectional_ring}). 

\section{Conclusion}

We first studied the effects of timestomping attacks on the age of gossip in a large fully connected network. We showed that one infected node in such a network can increase the age at all other nodes from $O(\log n)$ to $O(n)$ through timestamp manipulation. Further, we showed that the optimal behavior for the adversary is to reset the timestamps of all outgoing packets to current time thereby disguising them as current packets and of all incoming packets to an outdated time to prevent their acceptance at the infected node. We showed that if the adversary allows the infected node to accept even a very small fraction of the incoming packets from the network, then a large network can manage to curb the spread of stale files coming from the infected node and pull the network age back to $O(\log n)$. Additionally, we showed that if an infected node contacts only a single node instead of all nodes of the network, the system age can still be degraded to $O(n)$. We then analyzed the unidirectional ring network, where we showed that the adversarial effect on age scaling of a node is limited by its distance from the adversary, and the age scaling for a large fraction of the network continues to be $O(\sqrt{n})$, unchanged from the case with no adversary. We showed that the optimal behavior for the adversary is again to reset the timestamps of all outgoing packets to current time and of all incoming packets to an outdated time. Finally, we would like to point out that between these two network extremes representing very high and very low connectivity, there is a whole gamut of networks potentially exhibiting diverse age scaling behavior, which we plan to work on as future research direction.

\bibliographystyle{unsrt}
\bibliography{IEEEabrv,ref_priyanka}
\end{document}